\documentclass[aps,prd,preprint,superscriptaddress,nofootinbib]{revtex4-2}

\usepackage{supertabular} 
\usepackage{placeins}
\usepackage{epsfig}
\usepackage{graphicx}

\usepackage{float}
\usepackage{mathtools}
\usepackage{array}
\usepackage{dcolumn}

\usepackage{latexsym}
\usepackage{amsmath}
\usepackage{amssymb}
\usepackage{amsfonts}
\usepackage{bm}
\usepackage{physics}

\usepackage{color}
\definecolor{purple}{rgb}{0.5,0,0.5}
\definecolor{blue}{rgb}{0.0,0,0.9}
\definecolor{prdblue}{rgb}{0.133,0.118,0.498}
\usepackage[colorlinks=true, pdfstartview=FitV, linkcolor=prdblue, citecolor= prdblue, urlcolor=prdblue]{hyperref}

\begin{document}


\title{Detailed derivation of the $\mathbf{^3P_0}$ strong decay model applied to baryons}

\author{T.~Aguilar}
\email{telmo.aguilar@epn.edu.ec}
\affiliation{Departamento de Física, Escuela Politécnica Nacional, Quito 170143, Ecuador}

\author{A.~Capelo-Astudillo}
\affiliation{Departamento de Física, Escuela Politécnica Nacional, Quito 170143, Ecuador}

\author{M.~Conde-Correa}
\affiliation{Departamento de Física, Escuela Politécnica Nacional, Quito 170143, Ecuador}

\author{A.~Duenas-Vidal}
\email[]{alvaro.duenas@epn.edu.ec}
\affiliation{Departamento de Física, Escuela Politécnica Nacional, Quito 170143, Ecuador}

\author{P.~G.~Ortega}
\email[]{pgortega@usal.es}
\affiliation{Departamento de F\'isica Fundamental, Universidad de Salamanca, E-37008 Salamanca, Spain}
\affiliation{Instituto Universitario de F\'isica 
Fundamental y Matem\'aticas (IUFFyM), Universidad de Salamanca, E-37008 Salamanca, Spain}

\author{J.~Segovia}
\email[]{jsegovia@upo.es}
\affiliation{Departamento de Sistemas F\'isicos, Qu\'imicos y Naturales, Universidad Pablo de Olavide, E-41013 Sevilla, Spain}

\date{\today}

\begin{abstract}
We provide an in-detail derivation through the $^3P_0$ pair creation model of the transition matrix for a baryon decaying into a meson-baryon system.
The meson's analysis was conducted in Ref.~\cite{Segovia:2012cd} and we extend the same formalism to the baryon sector, focusing on the $\Delta(1232)\to \pi N$ strong decay width because all hadrons involved in the reaction are very well established, the two hadrons in the final state are stable, avoiding further analysis, all quarks are light and so equivalent, and the decay width of the process is relatively well measured. 
Taking advantage of a Gaussian expansion method for the hadron's radial wave functions, the expression of the invariant matrix element can be related with the mean-square radii of hadrons involved in the decay. We use their experimental measures in such a way that only the strength of the quark-antiquark pair creation from the vacuum is a free parameter. This is then taken from our previous study of strong decay widths in the meson sector, obtaining a quite compatible result with experiment for the calculated $\Delta(1232)\to \pi N$ decay width.
\end{abstract}

\maketitle



\section{Introduction}
\label{sec:introduction}

One of the main goals of nuclear and particle physics communities is the understanding of hadrons in terms of the elementary excitations of quantum chromo-dynamics (QCD), which are quarks and gluons (The interested reader is referred to the Particle Data Group and its topical mini-reviews~\cite{ParticleDataGroup:2022pth}). QCD is well understood in its high energy regime, where perturbative theoretical calculations has been contrasted with many experimental results since QCD's birth, 50 years ago; however, hadrons live in its non-perturbative regime where, \emph{a priori}, low-level rules produce high-level phenomena with enormous apparent complexity~\cite{Papavassiliou:2015aga}. That is to say, for instance, that less-than 2\% of a nucleon's mass can be attributed to the so-called current-quark masses that appear in QCD's Lagrangian, a phenomenon known as dynamical chiral symmetry breaking (DCSB). Another important non-perturbative effect is color confinement which basically states that quarks and gluons (color objects) are not those degrees-of-freedom readily accessible via experiment; \emph{i.e.}, they are confined inside hadrons. 

This complexity makes hadron spectroscopy, the collection of readily accessible  states constituted from gluons and quarks, the starting point for all further  investigations. A very successful classification scheme for hadrons in terms of  their valence quarks and antiquarks is the so-called quark model~\cite{Gell-Mann:1964ewy, Zweig:1964CERN}, which basically separates hadrons in quark-antiquark (meson) and three-quark (baryon) bound-states. The quark model, and its more modern variations and extensions, have received experimental verification beginning in the late 1960s and, despite some caveats, they have been demonstrated to be very valuable. For instance, the phenomenological quark models represent a reliable theoretical approach to hadron spectra in heavy quark sectors, are flexible enough in order to extend their applicability to exotic matter, and allow to compute easily electromagnetic, weak and strong reactions whose predictions have turned to be very useful for experimental searches.

Among the wide range of chiral quark models developed in the last 50 years~\cite{Fernandez:2021zjq}, our theoretical framework is a QCD-inspired constituent quark model (CQM) proposed in Ref.~\cite{Vijande:2004he} and extensively reviewed in Ref.~\cite{Segovia:2013wma}. Moreover, the CQM has been recently applied with success to conventional mesons containing heavy quarks, describing a wide range of physical observables that concern spectra~\cite{Segovia:2008zz, Segovia:2016xqb, Segovia:2015dia}, strong decays~\cite{Segovia:2009zz, Segovia:2012cd, Segovia:2013kg}, hadronic transitions~\cite{Segovia:2014mca, Segovia:2015raa, Martin-Gonzalez:2022qwd} as well as electromagnetic and weak reactions~\cite{Segovia:2011dg, Segovia:2011zza, Segovia:2012yh}. Besides, the interested reader could appreciate that the naive model has been extended to describe meson-meson molecules~\cite{Ortega:2012rs} and compact multiquark systems~\cite{Yang:2020atz}.

Trying to extend our CQM in the baryon sector, three steps must be taken: (i) the computation of baryon spectra, (ii) the modeling of a baryon decaying strongly into a meson plus another baryon and (iii) the description of baryon-meson interactions, and their resulting bound- and resonance-states, from the quark--(anti-)quark forces dictated by CQM. All of them are underway, see for example the advances done in the third case by one of us in, for instance, Refs.~\cite{Ortega:2012cx, Ortega:2014fha, Ortega:2014eoa}, but the first that has been completed by our group is the extension of the phenomenological ${}^3P_0$ model to the description of baryon strong decays. In fact, the same decay model was used in~\cite{Segovia:2012cd} to calculate the total strong decay widths of mesons which belong to heavy quark sectors. Therein, a global fit of the experimental data showed that, contrarily to the usual wisdom, the only free parameter of the ${}^3P_0$ model depends on the meson sector and thus the scale-dependent strength follows a logarithmic behavior with respect the typical scale of the particular meson sector (Eq.~(10) in Ref.~\cite{Segovia:2012cd}).

Hadron strong decay is a complex non-perturbative process that has not yet been
described from first principles of QCD. In the search for ways to explain it, Micu~\cite{Micu:1968mk} formulated the $^3P_0$ model in the 1960s as a way to obtain hadron's decay rates using the corresponding wave functions and a strength parameter as the only needed inputs. His approach was innovative for its simplicity and the few assumptions that were made. A few years later, Le Yaouanc~\cite{LeYaouanc:1972vsx} \emph{et al.} developed Micu's model using the work of Carlitz and Kislinger based on $SU(6)_w$ theory~\cite{Carlitz:1970xb}. Remarkable features of this work were the assumption that constituent quarks drive the decay process and the use of harmonic oscillator wave functions to find analytic expressions of the terms fitted by Micu from experiment. The only free parameter was then the so-called pair-creation constant, $\gamma$. The work of Le Yaouanc \emph{et al.} allowed to calculate many ratios between decay widths of mesons and baryons~\cite{LeYaouanc:1973ldf}, popularizing the model. In the following years, the $^3P_0$ model was widely used to describe decay properties of hadrons, such as charmonium states~\cite{LeYaouanc:1977fsz, LeYaouanc:1977gm}. In 1982 Hayne \emph{et al.} improved the analytic expression corresponding to the transition matrix~\cite{Hayne:1981zy}. In 1996, Blundell \emph{et al.} analyzed the data of various decay widths to fit the strength parameter $\gamma$~\cite{Blundell:1995ev}, finding a value which is frequently used in modern works~\cite{Guo:2019ytq}. In addition, parallel work in the flux-tube pair creation model~\cite{Kokoski:1985is} showed that it contains, and thus can be simplified to, the $^3P_0$ model making it even more famous. Recent variants of the $^{3}P_{0}$ model modify the pair production
vertex~\cite{Roberts:1997kq} or modulate the spatial dependence of the
pair-production amplitude~\cite{Chen:2017mug}.

This work consists on finding an analytical expression for the transition matrix of a baryon decaying into a meson-baryon system using the $^3P_0$ strong decay model to parametrize the needed quark-antiquark pair creation from the vacuum and Gaussian expansions of the hadron wave functions in order to simplify the evaluation of matrix elements, and express them in terms of the measured hadron sizes. The strength $\gamma$ of the decay interaction is fixed to our previous evaluation of meson strong decays in order to provide a free-parameter prediction of the $\Delta(1232)\to \pi N$ decay width; comparing it with the experimental value provide us an assessment of our calculation and the possible extension of our decay model from the meson sector to the baryon one.

This manuscript is organized as follows. After this introduction, Sec.~\ref{sec:theory} is devoted to a detailed description of the $^3P_0$ model and how to obtain the transition matrix, starting from the initial and final hadron states and the transition operator. A few assumptions are made to simplify the expression, the limitations of these are specified. Section~\ref{sec:results} provides an application of the model presented, obtaining the decay width of the process $\Delta(1232)\to \pi N$, specifying the data used. In this section, the quark-antiquark pair creation constant, $\gamma$, for baryons seems to follow the analytic expression presented in~\cite{Segovia:2012cd} for mesons. Finally, we summarize and draw some conclusions in Sec.~\ref{sec:summary}.


\begin{figure}[!t]
\centering
\includegraphics[scale=0.45]{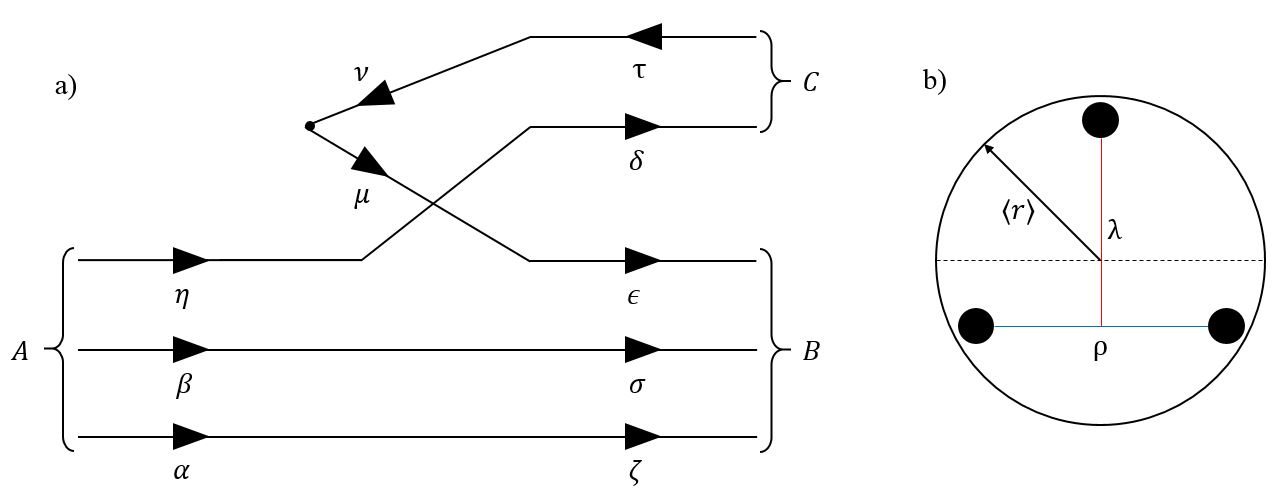}
\caption{\label{fig:General} \emph{Panel (a)} Feynman diagram for a baryon, A, decaying into a baryon, $B$, and a meson, $C$. \emph{Panel (b)} Schematic representation of a baryon as an sphere of radius $\langle r \rangle$; a particular set of Jacobi coordinates, $(\rho,\lambda)$, for the quarks in a baryon is also shown.}
\end{figure}

\section{The $\mathbf{^3P_0}$ model applied to baryons}
\label{sec:theory}

The quark-antiquark pair creation models consist on a phenomenological way to describe hadron strong decays. Among this kind of models, the so-called $^3P_0$ strong decay model is the most popular and basically states that the quark-antiquark pair, created from the vacuum, must conserve the vacuum's angular momentum, parity and charge conjugation, \emph{viz.} the quark-antiquark pair must have $J^{PC}=0^{++}$ quantum numbers. Another important property of the ${}^3P_0$ model is that it takes into account only diagrams in which the quark-antiquark pair separates into different final hadrons. This was originally motivated by experimental observations and it is known as the Okubo-Zweig-Iizuka(OZI)-rule~\cite{Okubo:1963fa, Zweig:1964jf, Iizuka:1966fk} which tells us that the disconnected diagrams are more suppressed than the connected ones.

The model defined as above describes baryon into meson$+$baryon strong decays as represented in panel (a) of Fig.~\ref{fig:General}. It thus has an associated transition operator given by
\begin{eqnarray}
T = -3\gamma'\sum_{\mu,\nu} \int d^3p_\mu d^3p_\nu \left[ \mathcal{Y}_1\left( \frac{\Vec{p}_\mu - \Vec{p}_{\nu}}{2}\right)  \otimes (s_\mu s_\nu)1 \right]_0 a^{\dagger}_{\mu}(\Vec{p}_\mu) b^{\dagger}_{\nu}(\Vec{p}_\nu)\delta^{(3)}(\Vec{p_\mu}+\Vec{p_\nu}),
\label{1Top}
\end{eqnarray}
where $\mu$ is the quark and $\nu$ is the antiquark created. The $3$-dimensional Dirac delta function, $\delta^{(3)}(\Vec{p_\mu}+\Vec{p_\nu})$, assures the conservation of momenta and the function $\mathcal{Y}_l(\Vec{p}\,) = p^l\, Y_{l}(\hat{p})$ is the solid harmonic that characterizes the angular momentum ($l=1$) of the pair created; one can also observe that it is coupled to the spin-$1$ of the pair in order to give total spin $J=0$. Meanwhile, $\gamma'$ is the only unknown constant of the ${}^3P_0$ model which characterizes the strength of the quark-antiquark pair creation from the vacuum and it is normally fitted to the data. Besides, it is important to note that this transition operator is a non-relativistic reduction of an interacting Hamiltonian involving Dirac quark fields that describes the production process~\cite{Segovia:2012cd}; observe therein that the $\sqrt{3}$ is replaced by $3$ when going from meson decays to baryon ones, since the term must cancel out with the color contribution.

The decay width of the process can be calculated using the following relation:
\begin{equation}
\Gamma_{A \rightarrow BC} = 2\pi \frac{E_B(k_0)E_C(k_0)}{m_Ak_0}\sum_{J_{BC},l}|\mathcal{M}_{A\rightarrow BC}|^2,
\end{equation} 
where $k_0$ is the relative momentum of the final products with respect to the initial state~\cite{Capstick:1992th}. The squared modulus of the invariant matrix element must be summed over all possible values of $J_{BC}$ and relative angular momentum $l$ whose inner product is equal to the total angular momentum of the decaying baryon, $J_A$.

In order to calculate the invariant matrix element that appears in the formula of the decay width,
\begin{equation}
\mathcal{M}_{A \rightarrow BC} = \delta^{(3)}(\Vec{K}_0) \langle BC | T | A \rangle \,,
\end{equation}
where $\Vec{K}_0$ is the center-of-mass momentum of the decaying baryon, one needs to establish expressions for the initial and final states:
\begin{align}
|A\rangle &= \int d^3p_\alpha d^3_\beta d^3p_\eta \delta^{(3)}(\Vec{P}_A - \Vec{K}_A)\chi_A C_A\phi_A(\Vec{p}_\alpha,\Vec{p}_\beta,\Vec{p}_\eta) a^\dagger_\alpha(\Vec{p}_\alpha) a^\dagger_\beta(\Vec{p}_\beta) a^\dagger_\eta(\Vec{p}_\eta) |0\rangle \,, \label{eq:A}\\
|BC\rangle &= \int d^3K_B d^3K_C \sum_{m,M_{BC}} \langle J_{BC}M_{BC}lm|J_AM_A \rangle \delta^{(3)}(\Vec{K}-\Vec{K}_0)\delta(k-k_0) \nonumber \\
&
\times \frac{Y_{lm}(\hat{k})}{k}\sum_{M_B,M_C,M_{I_B},M_{I_C}} \langle J_BM_BJ_CM_C|J_{BC}M_{BC} \rangle \langle I_BM_{I_B}I_CM_{I_C}|I_{BC}M_{I_{BC}} \rangle \nonumber \\
&
\times \int d^3p_\delta d^3p_\epsilon  d^3p_\zeta d^3p_\sigma d^3p_\tau \delta^{(3)}(\Vec{K}_B - \Vec{P}_B)\delta^{(3)}(\Vec{K}_C - \Vec{P}_C) \nonumber \\
&
\times \chi_B C_B\phi_B(\Vec{p}_\sigma,\Vec{p}_\zeta,\Vec{p}_\epsilon) a^\dagger_\sigma(\Vec{p}_\sigma) a^\dagger_\zeta(\Vec{p}_\zeta) a^\dagger_\epsilon(\Vec{p}_\epsilon) \chi_C C_C\phi_C(\Vec{p}_\delta,\Vec{p}_\tau) a^\dagger_\delta(\Vec{p}_\delta) b^\dagger_\tau(\Vec{p}_\tau) |0\rangle. \label{eq:BC}
\end{align}
In the equations above, the functions $\phi_{A,B,C}$ are the Fourier transforms of the hadron's wave functions in coordinate space, these describe the probability of finding the hadron in momentum space; moreover, the $C_{A,B,C}$ and $\chi_{A,B,C}$ are, respectively, the color and spin functions of the hadrons involved in the decay. In the final state $|BC\rangle$, the internal products assure the conservation of angular momentum and isospin between the baryon and meson in the final state, but also its coupling with the initial baryon state.

The invariant matrix element, ${\cal M}_{A\to BC}$, is a product of a color factor, a flavor factor and a spin-space overlap integral, \emph{i.e.}
\begin{equation}
\mathcal{M}_{A\rightarrow BC} = \mathcal{I}_{Color} \, \mathcal{I}_{Flavor} \, \mathcal{I}_{Spin-space} \,,
\end{equation}
in such a way that each component can be calculated separately.

\subsection{The spin-space contribution}

Before discussing the spin-space contribution, one needs to disentangle how many equivalent Feynman diagrams contribute to the same process, \emph{i.e.} the symmetry factor. Focusing on the ladder operators that appear in Eqs.~\eqref{1Top},~\eqref{eq:A} and~\eqref{eq:BC}, where combined adequately, we arrive at the following expression:
\begin{align}
\langle 0 | b_\tau(\Vec{p}_\tau) a_\delta(\Vec{p}_\delta) a_\epsilon(\Vec{p}_\epsilon) a_\zeta(\Vec{p}_\zeta) a_\sigma(\Vec{p}_\sigma) a^\dagger_\mu(\Vec{p}_\mu) b_\nu^\dagger(\Vec{p}_\nu) a_\alpha^\dagger(\Vec{p}_\alpha) a_\beta^\dagger(\Vec{p}_\beta) a_\eta^\dagger(\Vec{p}_\eta) |0\rangle \,.
\label{eq:ladder}
\end{align}
This product of creation and annihilation operators can be simplified. Since quarks are fermions, we use the anti-commutation relations of the ladder operators,
\begin{align}
\{ a_r(\Vec{p}\,),a^{\dagger}_s(\Vec{p}\,') \} &= a_r(\Vec{p}\,)a^{\dagger}_s(\Vec{p}\,')+a^{\dagger}_s(\Vec{p}\,')a_r(\Vec{p}\,)=\delta_{rs}\delta^{(3)}(\Vec{p} - \Vec{p}\,') \,, \\
\{ a_r(\Vec{p}\,),b^{\dagger}_s(\Vec{p}\,') \} &= \{ a_r(\Vec{p}\,),b_s(\Vec{p}\,')  \} = \{ a_r^{\dagger}(\Vec{p}\,),b_s(\Vec{p}\,')  \} = \{ a_r^{\dagger}(\Vec{p}\,),b^{\dagger}_s(\Vec{p}\,')  \} = 0 \,, \\
\{ a_r(\Vec{p}\,),a_s(\Vec{p}\,')  \} &= \{ a_r^{\dagger}(\Vec{p}\,),a^{\dagger}_s(\Vec{p}\,')  \} = \{ b_r(\Vec{p}\,),b_s(\Vec{p}\,')  \} = \{ b_r^{\dagger}(\Vec{p}\,),b^{\dagger}_s(\Vec{p}\,')  \} = 0 \,,
\end{align}
and arrange them in normal ordering to arrive at
\begin{eqnarray}
&&
\hspace{-0.80cm} \langle 0 | b_\tau a_\sigma a_\zeta a_\epsilon a_\delta a^\dagger_\mu b^\dagger_\nu a^\dagger_\alpha a^\dagger_\beta a^\dagger_\eta | 0 \rangle = \nonumber\\
		= && \delta_{\tau\nu} \delta_{\delta\mu} \delta_{\epsilon\alpha} \delta_{\zeta\beta} \delta_{\sigma\eta} - \delta_{\tau\nu} \delta_{\delta\mu} \delta_{\epsilon\alpha} \delta_{\sigma\beta} \delta_{\zeta\eta}  - \delta_{\tau\nu} \delta_{\delta\mu} \delta_{\zeta\alpha} \delta_{\epsilon\beta} \delta_{\sigma\eta} - \delta_{\tau\nu}\delta_{\epsilon \mu} \delta_{\delta\alpha} \delta_{\zeta\beta} \delta_{\sigma\eta} \nonumber\\
        + &&\delta_{\tau\nu} \delta_{\delta\mu} \delta_{\zeta\alpha} \delta_{\sigma\beta} \delta_{\epsilon\eta} + \delta_{\tau\nu} \delta_{\delta\mu} \delta_{\sigma\alpha} \delta_{\epsilon\beta} \delta_{\zeta\eta}  + \delta_{\tau\nu}\delta_{\epsilon \mu} \delta_{\delta\alpha} \delta_{\sigma\beta} \delta_{\zeta\eta} + \delta_{\tau\nu}\delta_{\epsilon \mu} \delta_{\zeta\alpha} \delta_{\delta\beta} \delta_{\sigma\eta} \nonumber\\
        + &&\delta_{\tau\nu} \delta_{\zeta\mu} \delta_{\delta\alpha} \delta_{\epsilon\beta} \delta_{\sigma\eta} - \delta_{\tau\nu} \delta_{\delta\mu} \delta_{\sigma\alpha} \delta_{\zeta\beta} \delta_{\epsilon\eta} - \delta_{\tau\nu}\delta_{\epsilon \mu} \delta_{\zeta\alpha} \delta_{\sigma\beta} \delta_{\delta\eta} - \delta_{\tau\nu}\delta_{\epsilon \mu} \delta_{\sigma\alpha} \delta_{\delta\beta} \delta_{\zeta\eta} \nonumber\\
        - &&\delta_{\tau\nu} \delta_{\zeta\mu} \delta_{\delta\alpha} \delta_{\sigma\beta} \delta_{\epsilon\eta} - \delta_{\tau\nu}\delta_{\zeta\mu} \delta_{\epsilon\alpha} \delta_{\delta\beta} \delta_{\sigma\eta} - \delta_{\tau\nu} \delta_{\sigma\mu} \delta_{\delta\alpha} \delta_{\epsilon\beta} \delta_{\zeta\eta} + \delta_{\tau\nu}\delta_{\epsilon \mu} \delta_{\sigma\alpha} \delta_{\zeta\beta} \delta_{\delta\eta} \nonumber\\
        + &&\delta_{\tau\nu}\delta_{\zeta\mu} \delta_{\epsilon\alpha} \delta_{\sigma\beta} \delta_{\delta\eta} + \delta_{\tau\nu} \delta_{\zeta\mu} \delta_{\sigma\alpha} \delta_{\delta\beta} \delta_{\epsilon\eta} + \delta_{\tau\nu} \delta_{\sigma\mu} \delta_{\delta\alpha} \delta_{\zeta\beta} \delta_{\epsilon\eta} + \delta_{\tau\nu} \delta_{\sigma\mu} \delta_{\epsilon\alpha} \delta_{\delta\beta} \delta_{\zeta\eta} \nonumber\\
        - &&\delta_{\tau\nu} \delta_{\zeta\mu} \delta_{\sigma\alpha} \delta_{\epsilon\beta} \delta_{\delta\eta} - \delta_{\tau\nu} \delta_{\sigma\mu} \delta_{\epsilon\alpha} \delta_{\zeta\beta} \delta_{\delta\eta} - \delta_{\tau\nu} \delta_{\sigma\mu} \delta_{\zeta\alpha} \delta_{\delta\beta} \delta_{\epsilon\eta} + \delta_{\tau\nu} \delta_{\sigma\mu} \delta_{\zeta\alpha} \delta_{\epsilon\beta} \delta_{\delta\eta}.
\end{eqnarray}
Note here that we have done an abuse of notation, $\delta_{ab}\,\delta^{(3)}(\vec{p}_a-\vec{p}_b) \equiv \delta_{ab}$.

\begin{figure}[t!]
    \centering
    \begin{tabular}{c c c}
        \begin{minipage}{.3\textwidth}
        \includegraphics[scale=0.27]{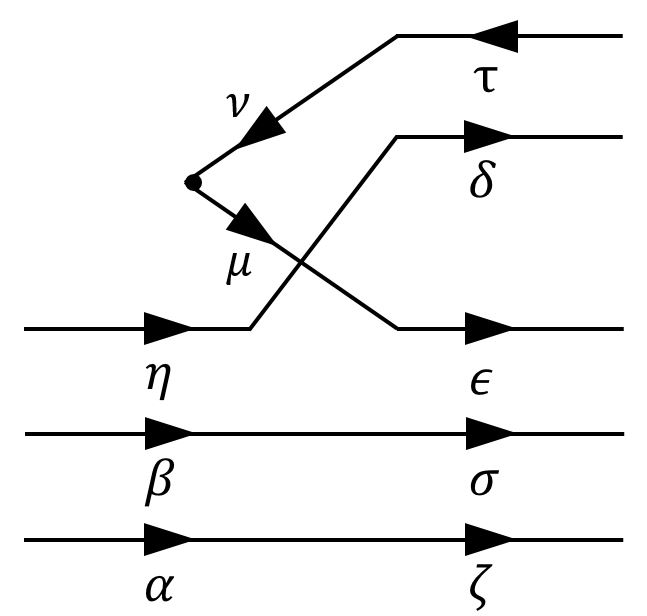}
        \end{minipage} 
        & \begin{minipage}{.3\textwidth}
        \includegraphics[scale=0.27]{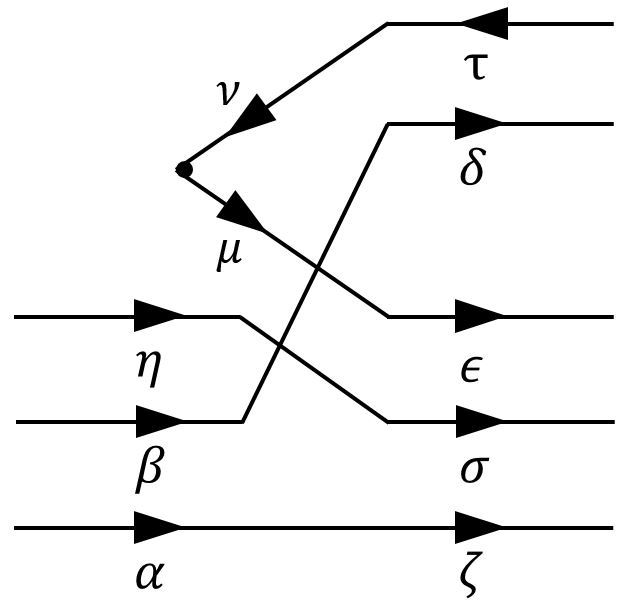}
        \end{minipage} & 
        \begin{minipage}{.3\textwidth}
        \includegraphics[scale=0.27]{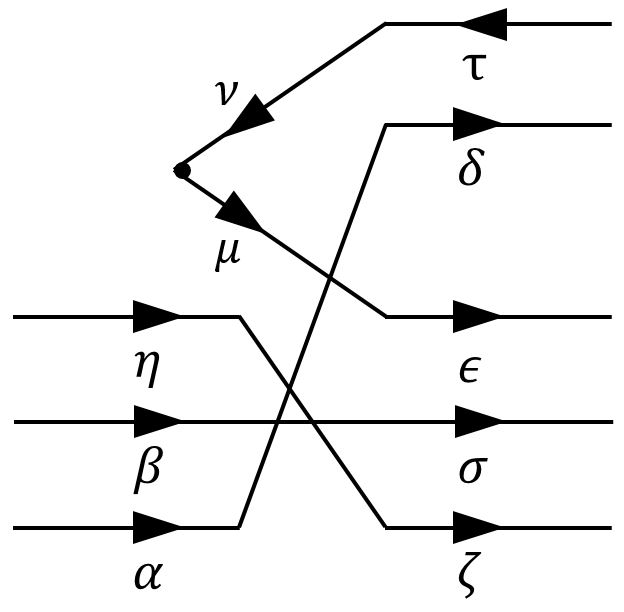}
        \end{minipage}\\
         $d_{\eta\epsilon}$ & $d_{\beta\epsilon}$ & $d_{\alpha\epsilon}$\\
         &  & \\
         \begin{minipage}{.3\textwidth}
        \includegraphics[scale=0.27]{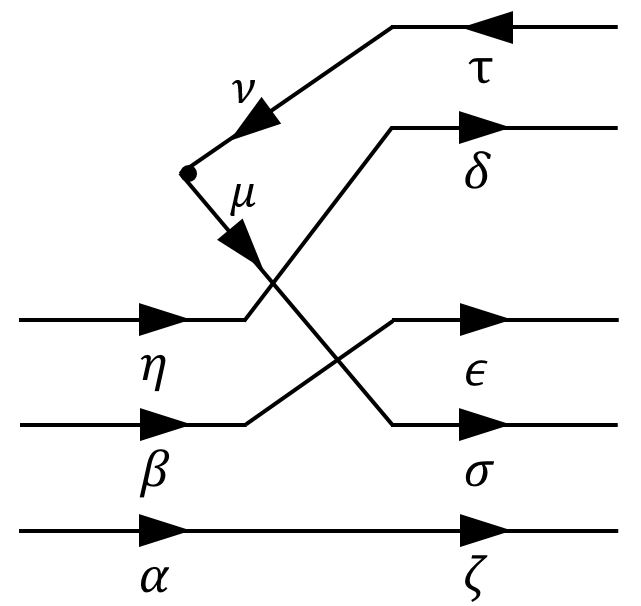}
        \end{minipage} 
        & \begin{minipage}{.3\textwidth}
        \includegraphics[scale=0.27]{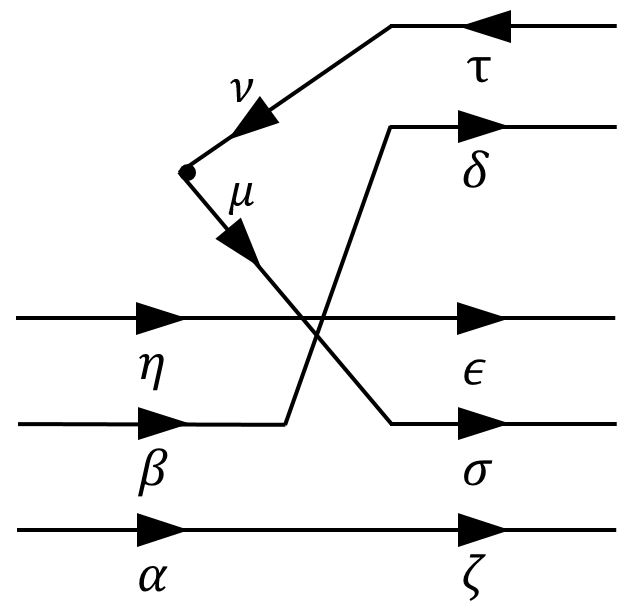}
        \end{minipage} & 
        \begin{minipage}{.3\textwidth}
        \includegraphics[scale=0.27]{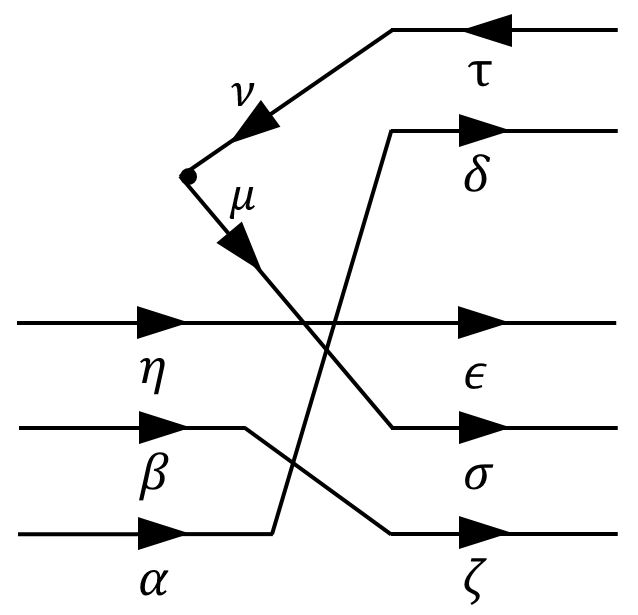}
        \end{minipage}\\
         $d_{\eta\sigma}$ & $d_{\beta\sigma}$ & $d_{\alpha\sigma}$\\
         &  & \\
         \begin{minipage}{.3\textwidth}
        \includegraphics[scale=0.27]{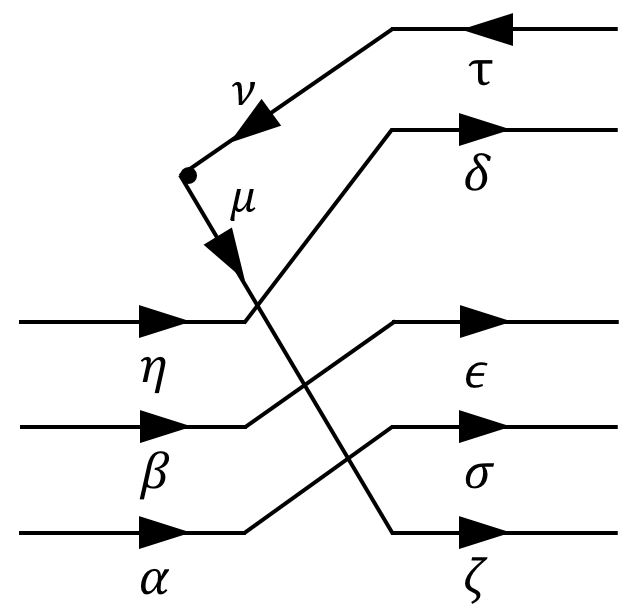}
        \end{minipage} 
        & \begin{minipage}{.3\textwidth}
        \includegraphics[scale=0.27]{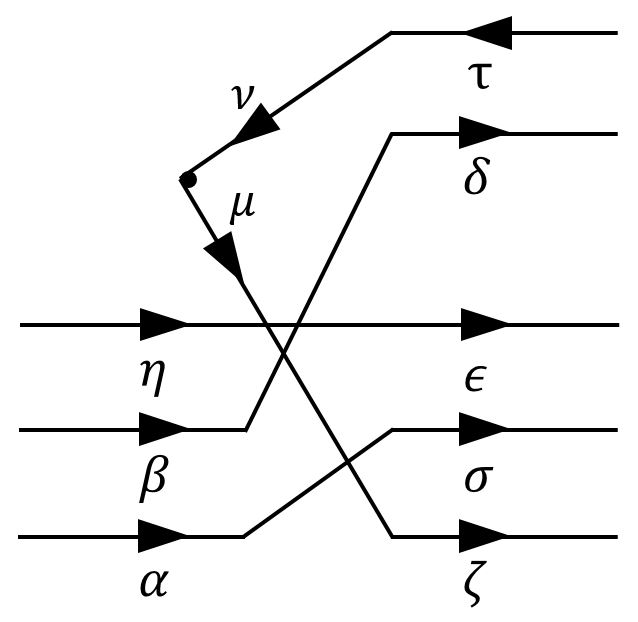}
        \end{minipage} & 
        \begin{minipage}{.3\textwidth}
        \includegraphics[scale=0.27]{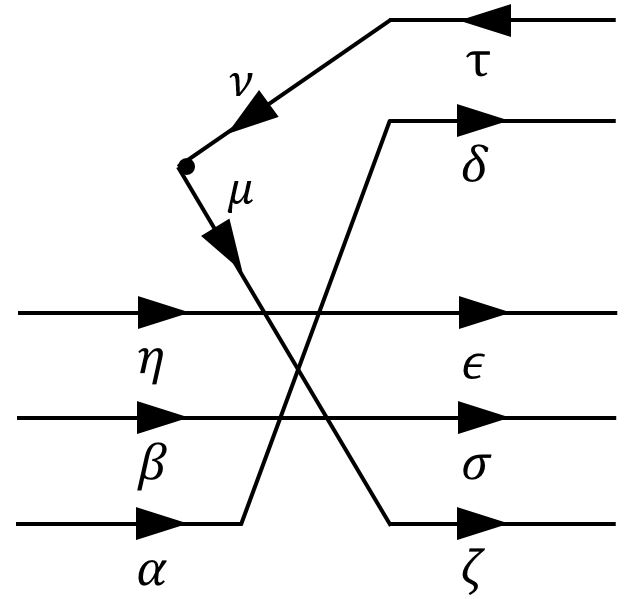}
        \end{minipage}\\
         $d_{\eta\zeta}$ & $d_{\beta\zeta}$ & $d_{\alpha\zeta}$\\
         &  & \\
    \end{tabular}
    \caption{Decay process for a Baryon. Below each diagram the notation $d_{ab}$ characterises the diagram using the quark $(a)$ that begins in the baryon and end in the meson and the quark $(b)$ that comes form the pair created and end in the baryon.}
    \label{fig:CompDiag}
\end{figure}

Each term is a different process that can be represented by a Feynman diagram. Following the OZI-rule, those terms with the factor $\delta_{\delta\mu}$ can be eliminated, the remaining ones are pictorially shown in Fig.~\ref{fig:CompDiag}. If all quarks and antiquarks involved in the baryon strong decay are indistinguishable, the diagrams can be taken as equivalent. Therefore, the final result may be written as
\begin{equation}
\langle 0|b_\tau a_\sigma a_\zeta a_\epsilon a_\delta a^\dagger_\mu b^\dagger_\nu a^\dagger_\alpha a^\dagger_\beta a^\dagger_\eta|0 \rangle = - 18 \, \delta_{\tau\nu} \delta_{\epsilon\mu} \delta_{\zeta\alpha} \delta_{\sigma\beta} \delta_{\delta\eta} \,.
\label{Equiva}
\end{equation}
If this equivalence between quarks and antiquarks do not hold, the contribution of different diagrams differs between them but do not change the subsequent expressions significantly and thus one may straightforwardly extend the computation to be described below.

Now, the spin-space contribution can be separated in two terms, one collects the coupling of angular momentum and spin, $(\mathcal{J})$, and the other deals with linear momenta, $(\mathcal{E})$,
\begin{equation}
\mathcal{I}_{Spin-space} = 54\gamma' \sum_{L_{BC},L,S} \mathcal{J}(A\rightarrow BC) \mathcal{E}(A\rightarrow BC) \,.
\end{equation}

The hadron's total angular momentum, $J$, represents a coupling between its angular momentum, $L$, and its spin, $S$. Therefore, the initial form of $\mathcal{J}(A\rightarrow BC)$ is
\begin{equation}
\mathcal{J}(A\rightarrow BC) = \left\{[L_BS_B]_{J_B} [L_CS_C]_{J_C} \right\}_{J_{BC}}^* \{[L_{BC}S]_{J_{BC}} \, l\}_{J_A}^* \{[L_AS_A]_{J_A} [11]_0\}_{J_A} \,,
\label{eq:JJ}
\end{equation}
where the extra $[11]_0$ is added to take into account the quantum numbers of the quark-antiquark pair created from the vacuum. Note also that complex conjugate symbols affect to final states as expected.

Equation~\eqref{eq:JJ} is a matrix element written in terms of hadron's individual $L-S$ coupling into $J$. The final expression must have a total angular momentum $(L)$, inner sum of all the angular momenta of the particles, and a total spin $(S)$, inner sum of all the spins of the particles. These final angular momentum and spin must be then coupled to the total angular momentum of the decaying baryon. These transformations can be done using Wigner symbols~\cite{QTAM}:
\begin{equation}
\{ [L_1S_1]_{J_1},[L_2S_2]_{J_2} \}_{J_T} = 
\Pi_{J_1,J_2,L_T,S_T} 
\begin{Bmatrix}
L_1 && L_2 && L_T\\
S_1 && S_2 && S_T\\
J_1 && J_2 && J_T
\end{Bmatrix} \{[L_1L_2]_{L_T},[S_1S_2]_{S_T}] \}_{J_T} ,
\end{equation}
where $\Pi_l=\sqrt{2l+1}$ is used to simplify the notation. With this relation the couplings of the initial state and the pair created can be changed as
\begin{equation}
    \{[L_AS_A]_{J_A} [11]_0\}_{J_A}  = \sum_{L,S} \Pi_{L,S,J_A,0} \begin{Bmatrix}
    L_A && S_A && J_A\\
    1 && 1 && 0\\
    L && S && J_A
    \end{Bmatrix} \left\{[L_A1]_L [S_A1]_S \right\}_{J_A} \,.
\end{equation}
Since the $9J$-symbol contain a zero in one of its components, it can be reduced to a $6J$-symbol~\cite{QTAM}:
\begin{equation}
\left\{[L_AS_A]_{J_A} [11]_0 \right\}_{J_A} = \sum_{L,S} (-1)^{S + J_A + L_A + 1} \frac{\Pi_{L,S}}{\sqrt{3}}\begin{Bmatrix}
L && S && J_A\\
S_A && L_A && 1
\end{Bmatrix} \left\{[L_A1]_L [S_A1]_S \right\}_{J_A}.
\end{equation}
A similar transformation can be done for the final state,
\begin{equation}
\left\{[L_BS_B]_{J_B} [L_CS_C]_{J_C} \right\}_{J_{BC}}^* = 
\sum_{L_{BC},S} \Pi_{L_{BC},S,J_{B},J_C} \begin{Bmatrix}
L_B && S_B && J_B\\
L_C && S_C && J_C\\
L_{BC} && S && J_{BC}
\end{Bmatrix} \left\{[L_BL_C]_{L_{BC}} [S_BS_C]_S \right\}_{J_A}^*
\end{equation}
where the conservation of spin is used, simplifying $S_{BC}=S$. The baryon and meson in the final state have a relative angular momentum between them denoted by $l$; reordering terms as indicated in~\cite{QTAM}, we arrive at:
\begin{align}
&
\{[L_{BC} S]_{J_{BC}} \, l\}_{J_A}^* = (-1)^{L_{BC}+S-J_{BC}} \{[SL_{BC}]_{J_{BC}} \, l\}_{J_A}^* \nonumber\\
&
= (-1)^{L_{BC}+S-J_{BC}} \sum_{L} (-1)^{L_{BC}+S+J_A+l} \Pi_{L,J_{BC}} \begin{Bmatrix}
S && L_{BC} && J_{BC}\\
l && J_A && L
\end{Bmatrix} \{S [L_{BC}\,l]_L\}_{J_A}^* \nonumber\\
&
= \sum_{L} (-1)^{2L_{BC}+2S+J_A+l-J_{BC}} \Pi_{L,J_{BC}} \begin{Bmatrix}
S && L_{BC} && J_{BC}\\
l && J_A && L
\end{Bmatrix} (-1)^{S+L-J_A} \{[L_{BC}\,l]_L S\}_{J_A}^* \nonumber\\
&
= \sum_{L} (-1)^{S+L+l-J_{BC}} \Pi_{L,J_{BC}} \begin{Bmatrix}
S && L_{BC} && J_{BC}\\
l && J_A && L
\end{Bmatrix} \{[L_{BC}\,l]_L S\}_{J_A}^* \,.
\end{align}

The spin couplings can be also simplified in the following; the corresponding matrix element,
\begin{equation}
\{[s_\mu s_\beta s_\alpha]_{S_B}[s_\nu s_\eta]_{S_C}\}_S^* \{[s_\alpha s_\beta s_\eta]_{S_A}[s_\mu s_\nu]_{1}\}_S \,,
\end{equation}
has been written taking into account the delta-functions of Eq.~\eqref{Equiva}. Now, because the couplings are binary operations, the spins of the quarks inside baryons must be ordered. Using the Jacobi coordinate system $(\rho,\lambda)$ shown in panel (b) of Fig.~\ref{fig:CompDiag}, the named $\rho$-spin can be introduced,
\begin{equation}
s_\rho = s_\alpha \otimes s_\beta \,,
\end{equation}
as the spin of the non-interacting quarks during the decay, \emph{i.e.} those quarks that do not change their properties in the process. Then, the spin of the baryons as follows,
\begin{align}
\left[ s_\alpha s_\beta s_\eta \right]_{S_A} &= \left[s_\rho s_\eta \right]_{S_A} \,, \nonumber \\
\left[ s_\mu s_\beta s_\alpha \right]_{S_B} &= \left[ s_\mu \left( s_\beta s_\alpha \right)_{s_\rho} \right]_{S_B} = (-1)^{s_\alpha + s_\beta + s_\mu - S_B} \left[ s_\rho s_\mu \right]_{S_B} \,,
\end{align}
and so the spin conservation can be expressed using a $9J$-symbol,
\begin{align}
&
\{[s_\mu s_\beta s_\alpha]_{S_B} [s_\nu s_\eta]_{S_C}\}_S^* \{[s_\alpha s_\beta s_\eta]_{S_A}[s_\mu s_\nu]_{1}\}_S = \nonumber\\
&
=(-1)^{s_\alpha + s_\beta + s_\mu + s_\nu + s_\eta - S_B - S_C} \Pi_{S_B,S_C,S_A,1} \begin{Bmatrix}
        s_\rho && s_\mu && S_B\\
        s_\eta && s_\nu && S_C\\
        S_A && 1 && S
    \end{Bmatrix}.
\end{align}
Once all couplings are modified, the final expression for $\mathcal{J}(A\rightarrow BC)$ looks like
\begin{align}
\mathcal{J}(A\rightarrow BC) &= (-1)^{3/2 - S_B - S_C + L_A + L + l + J_A - J_{BC}} \, \Pi_{L_{BC},L,L,J_B,J_C,J_{BC},S_A,S_B,S_C,S,S} \nonumber\\
&
\times \begin{Bmatrix}
    L && S && J_A\\
    S_A && L_A && 1
    \end{Bmatrix} \begin{Bmatrix}
        S && L_{BC} && J_{BC}\\
        l && J_A && L
    \end{Bmatrix} \nonumber\\
&
\times \begin{Bmatrix}
        s_\rho && 1/2 && S_B\\
        1/2 && 1/2 && S_C\\
        S_A && 1 && S
    \end{Bmatrix} \begin{Bmatrix}
    L_B && S_B && J_B\\
    L_C && S_C && J_C\\
    L_{BC} && S && J_{BC}
    \end{Bmatrix}.
\end{align}

The remaining term to be calculated is the linear momentum contribution, whose initial expression is
\begin{align}
&
\mathcal{E}(A \rightarrow BC) = \int d^3K_B d^3K_C d^3p_\alpha d^3 p_\beta d^3p_\eta d^3p_\mu d^3p_\nu \nonumber\\
&
\times \delta^{(3)}(\Vec{K} - \Vec{K}_0) \delta^{(3)}(\Vec{K}_A - \Vec{P}_A) \delta^{(3)}(\Vec{K}_B - \Vec{P}_B) \delta^{(3)}(\Vec{K}_C - \Vec{P}_C) \delta^{(3)}(\Vec{p}_\mu + \Vec{p}_\nu) \frac{\delta(k-k_0)}{k} \nonumber\\
&
\times \left\{ \left[ \phi_B(\Vec{p}_\sigma,\Vec{p}_\zeta,\Vec{p}_\epsilon) \phi_C(\Vec{p}_\delta,\Vec{p}_\tau) \right]_{L_{BC}} Y_l(\hat{k}) \right\}_L^* \left\{ \phi_A(\Vec{p}_\alpha,\Vec{p}_\beta,\Vec{p}_\eta) \mathcal{Y}_1 \left( \frac{\Vec{p}_\mu - \Vec{p}_\nu}{2} \right) \right\}_L \,.
\end{align}
This expression can be simplified defining a new set of coordinates:
\begin{equation}
\begin{matrix*}[l]
        \Vec{P_A} = \Vec{p}_\alpha + \Vec{p}_\beta + \Vec{p}_\eta \,, && \quad\quad && \Vec{P}_C = \Vec{p}_\delta + \Vec{p}_\tau \,, \\
        \Vec{p}_{\rho_A} = \frac{\omega_\beta \Vec{p}_\alpha - \omega_\alpha \Vec{p}_\beta}{\omega_{\alpha\beta}} \,, && \quad && \Vec{p}_C = \frac{\omega_\delta\Vec{p}_\tau - \omega_\tau \Vec{p}_\delta}{\omega_{\delta\tau}} \,, \\
        \Vec{p}_{\lambda_A} = \frac{\omega_\eta(\Vec{p}_\alpha + \Vec{p}_\beta)- \omega_{\alpha\beta}\Vec{p}_\eta}{\omega_{\alpha\beta\eta}} \,, && \quad\quad && \Vec{P} = \Vec{p}_\mu + \Vec{p}_\nu \,, \\
        \Vec{P_B} = \Vec{p}_\zeta + \Vec{p}_\sigma + \Vec{p}_\epsilon \,, && \quad\quad && \Vec{p} = \frac{\Vec{p}_\mu - \Vec{p}_\nu}{2} \,, \\
        \Vec{p}_{\rho_B} = \frac{\omega_\sigma \Vec{p}_\zeta - \omega_\zeta \Vec{p}_\sigma}{\omega_{\zeta\sigma}} \,, && \quad\quad && \Vec{K} = \Vec{K}_B + \Vec{K}_C \,, \\
        \Vec{p}_{\lambda_B} = \frac{\omega_\epsilon(\Vec{p}_\zeta + \Vec{p}_\sigma)- \omega_{\zeta\sigma}\Vec{p}_\epsilon}{\omega_{\zeta\sigma\epsilon}} \,, && \quad\quad && \Vec{k} = \frac{\omega_C\Vec{K}_B - \omega_B\Vec{K}_C}{\omega_{BC}} \,,
    \end{matrix*}
\end{equation}
where we have introduce a so-called reduced mass convention which redefines all masses in terms of a reference one, $m$,
\begin{equation}
\begin{matrix}
\omega_\alpha = \frac{m_\alpha}{m} \,, && \quad\quad && \omega_{\alpha\beta} = \omega_\alpha + \omega_\beta \,.
\end{matrix}
\end{equation}
The delta functions related with momenta provide an additional set of conditions, 
\begin{align}
&\Vec{K} = \Vec{K}_0 = \Vec{K}_A = \Vec{P}_A = 0 \,,\nonumber\\
&\Vec{K}_B = \Vec{P}_B \,,\nonumber\\
&\Vec{K}_C = \Vec{P}_C \,,\nonumber\\
&\Vec{p}_\mu + \Vec{p}_\nu = \Vec{P} =0 \,,
\end{align}
where it is important to note that the center-of-mass of baryon $A$ is taken as the center of mass of the interaction. Now, the equivalences in momenta eliminate some integrals and the reaming variables can be written in terms of the following ones:
\begin{align}
\Vec{p} &= \Vec{p}_\mu = -\Vec{p}_\nu \,,\nonumber\\
\Vec{k} &= \Vec{p}_{\lambda_A} + \Vec{p} \,,\nonumber\\
\Vec{p}_\rho &= \Vec{p}_{\rho_A} = \Vec{p}_{\rho_B} \,.
\end{align}
Then, the simplified expression for $\mathcal{E}(A \rightarrow BC)$ is
\begin{align}
\mathcal{E}(A \rightarrow BC) &= \int d^3p d^3k d^3p_\rho\frac{\delta(k-k_0)}{k^{l+1}} \nonumber\\
&
\times \left\{ \left[ \phi_B(\Vec{p}_\sigma,\Vec{p}_\zeta,\Vec{p}_\epsilon) \phi_C(\Vec{p}_\delta,\Vec{p}_\tau) \right]_{L_{BC}} \mathcal{Y}_l(\hat{k}) \right\}_L^* \left\{ \phi_A(\Vec{p}_\alpha,\Vec{p}_\beta,\Vec{p}_\eta) \mathcal{Y}_1 \left( \frac{\Vec{p}_\mu - \Vec{p}_\nu}{2} \right) \right\}_L.
\label{eq:E1}
\end{align}

Continuing with the calculation, the hadron wave functions can be separated in radial and angular parts,
\begin{align}
\phi_A(\Vec{p}_\alpha,\Vec{p}_\beta,\Vec{p}_\eta) &= f_{\lambda_A}(\Vec{p}_{\lambda_A}) f_{\rho}(\Vec{p}_{\rho}) \left[\mathcal{Y}_{L_{\lambda_A}}(\Vec{p}_{\lambda_A}) \mathcal{Y}_{L_{\rho}}(\Vec{p}_{\rho})\right]_{L_A} \,, \\
\phi_B(\Vec{p}_\sigma,\Vec{p}_\zeta,\Vec{p}_\epsilon) &= f_{\lambda_B}(\Vec{p}_{\lambda_B}) f_{\rho}(\Vec{p}_{\rho}) \left[\mathcal{Y}_{L_{\lambda_B}}(\Vec{p}_{\lambda_B}) \mathcal{Y}_{L_{\rho}}(\Vec{p}_{\rho})\right]_{L_B} \,, \\
\phi_C(\Vec{p}_C) &= f_C(\Vec{p}_C)\mathcal{Y}_{L_C}(\Vec{p}_C) \,,
\end{align}
where the solid spherical harmonics take into account the Jacobi coordinate decomposition of a baryon system and the radial parts are assumed to be Gaussian functions,
\begin{align}
    f_{\lambda_A}(\Vec{p}_{\lambda_A}) &= \sum_i d_i^{\lambda_A} \exp\left( -\frac{{\lambda_A}_i}{2} {p}_{\lambda_A}^2 \right) \,, \\
    f_{\lambda_B}(\Vec{p}_{\lambda_B}) &= \sum_j d_j^{\lambda_B} \exp\left( -\frac{{\lambda_B}_j}{2} {p}_{\lambda_B}^2 \right) \,, \\
    f_\rho(\Vec{p}_\rho) &= \sum_k d_k^\rho \exp\left( -\frac{\rho_k}{2} {p}_\rho^2 \right) \,, \\
    f_C(\Vec{p}_C) &= \sum_{l'} d_{l'}^C \exp\left( -\frac{C_{l'}}{2} {p}_C^2 \right) \,,
\end{align}
where the constants could be computed theoretically from hadron spectra or fitted to experimental data of hadron radii.\footnote{In order to test this theoretical calculation we are going to follow the second strategy and we leave the microscopic calculation of the wave functions for a later publication because we have not yet developed the program that would calculate eigenenergies and eigenfunctions of baryons.} The limits of the sums are fixed according to the precision required. Inserting the above expressions in Eq.~\eqref{eq:E1}, we arrive at
\begin{align}
&
\mathcal{E}(A_{ik} \rightarrow B_{jk}C_{l'}) = \sum_{ijkl'} d_i^{\lambda_A} d_j^{\lambda_B} (d_k^\rho)^2 d_{l'}^C \nonumber\\
&
\times \int d^3p d^3k d^3p_{\rho} \frac{\delta(k-k_0)}{k^{l+1}} \exp\left( -\frac{1}{2} [{\lambda_A}_i p_{\lambda_A}^2 + {\lambda_B}_j p_{\lambda_B}^2 + \rho_k p_\rho^2 + C_{l'} p_C^2] \right) \nonumber\\
&
\times \left\{ \left[ \left[\mathcal{Y}_{L_{\lambda_B}}(\Vec{p}_{\lambda_B}) \mathcal{Y}_{L_{\rho}}(\Vec{p}_{\rho})\right]_{L_B} \mathcal{Y}_{L_C}(\Vec{p}_C) \right]_{L_{BC}} \mathcal{Y}_l(\Vec{k}) \right\}_L^* \left\{ \left[\mathcal{Y}_{L_{\lambda_A}}(\Vec{p}_{\lambda_A}) \mathcal{Y}_{L_{\rho}}(\Vec{p}_{\rho})\right]_{L_A} \mathcal{Y}_1 \left( \Vec{p}\,\right) \right\}_L \,.
\label{eq:E2}
\end{align}
We need now that all functions of Eq.~\eqref{eq:E2} be expressed in terms of the integration variables. In order to do that we define $\Vec{q} = \Vec{p} - x\Vec{k}$, where $x$ could be any number; note also that $p$ and $p_\rho$ continue to be variables of the integral. Therefore,
\begin{align}
\Vec{p}_{\lambda_A} &= \left(1-x\right)\Vec{k} - \Vec{q} \,, \nonumber\\
\Vec{p}_{\lambda_B} &= \left(\frac{\omega_{\mu}}{\omega_{\alpha\beta\mu}} - x\right)\Vec{k} - \Vec{q} \,, \nonumber\\
\Vec{p}_C &= \left(\frac{\omega_\mu}{\omega_{\eta\mu}} - x \right) \Vec{k} - \Vec{q} \,.
\end{align}
The terms are squared,
\begin{align}
p_{\lambda_A}^2 &= \left(1-x\right)^2k^2 + q^2 - 2\left(1-x\right) \Vec{k}\cdot\Vec{q} \,, \nonumber \\
p_{\lambda_B}^2 &= \left(\frac{\omega_{\mu}}{\omega_{\alpha\beta\mu}} - x\right)^2k^2 + q^2 - 2\left(\frac{\omega_{\mu}}{\omega_{\alpha\beta\mu}} - x\right)\Vec{k}\cdot\Vec{q} \,, \nonumber \\
p_C^2 &= \left(\frac{\omega_\mu}{\omega_{\eta\mu}} - x \right)^2 k^2 + q^2 + 2\left(\frac{\omega_\mu}{\omega_{\eta\mu}} - x \right)\Vec{k}\cdot\Vec{q} \,,
\end{align}
and replaced in the exponential argument as
\begin{align}
{\lambda_A}_i p_{\lambda_A}^2 + {\lambda_B}_j p_{\lambda_B}^2 + C_l p_C^2 &=
k^2 \left[ {\lambda_A}_i(1-x)^2 + {\lambda_B}_j \left(\frac{\omega_{\mu}}{\omega_{\alpha\beta\mu}} - x\right)^2 + C_l \left(\frac{\omega_\mu}{\omega_{\eta\mu}} - x \right)^2\right] \nonumber\\
&
+ q^2 \left[ {\lambda_A}_i + {\lambda_B}_j + C_l \right] \nonumber\\
&
- 2\Vec{k}\cdot\Vec{q} \left[ {\lambda_A}_i (1-x) + {\lambda_B}_j \left(\frac{\omega_{\mu}}{\omega_{\alpha\beta\mu}} - x\right) + C_l \left(\frac{\omega_\mu}{\omega_{\eta\mu}} - x \right) \right] \,.
\end{align}
Now, in order to eliminate the $\Vec{k}\cdot\Vec{q}$ term, $x$ is fixed to the following value,
\begin{equation}
x = \frac{{\lambda_A}_i + {\lambda_B}_j\frac{\omega_{\mu}}{\omega_{\alpha\beta\mu}} + C_l\frac{\omega_\mu}{\omega_{\eta\mu}}}{{\lambda_A}_i + {\lambda_B}_j + C_l}.
\label{eq:C1}
\end{equation}
To simplify more the notation, the parameters
\begin{align}
A &= {\lambda_A}_i + {\lambda_B}_j\frac{\omega_{\mu}}{\omega_{\alpha\beta\mu}} + C_l\frac{\omega_\mu}{\omega_{\eta\mu}} \,, \label{eq:C2} \\
2B &= {\lambda_A}_i + {\lambda_B}_j + C_l \,, \label{eq:C3} \\
2D &= {\lambda_A}_i(1-x)^2 + {\lambda_B}_j \left(\frac{\omega_{\mu}}{\omega_{\alpha\beta\mu}} - x\right)^2 + C_l \left(\frac{\omega_\mu}{\omega_{\eta\mu}} - x \right)^2 \,, \label{eq:C4}
\end{align}
are defined, where $x=\frac{A}{2B}$. Therefore, the linear momentum contribution can be now written as
\begin{align}
&
\mathcal{E}(A_{ik} \rightarrow B_{jk}C_{l'}) = \sum_{ijkl'} d_i^{\lambda_A} d_j^{\lambda_B} (d_k^\rho)^2 d_{l'}^C \int d^3q d^3k d^3p_{\rho} \frac{\delta(k-k_0)}{k^{l+1}} \exp\left( - Bq^2 - Dk^2  -\frac{\rho_k}{2} p_\rho^2 \right) \nonumber\\
&
\times \left\{ \left[ \left[\mathcal{Y}_{L_{\lambda_B}}(\Vec{p}_{\lambda_B}) \mathcal{Y}_{L_{\rho}}(\Vec{p}_{\rho})\right]_{L_B} \mathcal{Y}_{L_C}(\Vec{p}_C) \right]_{L_{BC}} \mathcal{Y}_l(\Vec{k}) \right\}_L^* \left\{ \left[\mathcal{Y}_{L_{\lambda_A}}(\Vec{p}_{\lambda_A}) \mathcal{Y}_{L_{\rho}}(\Vec{p}_{\rho})\right]_{L_A} \mathcal{Y}_1 \left( \Vec{p}\, \right) \right\}_L \,,
\label{eq:E3}
\end{align}
but the second line of Eq.~\eqref{eq:E3} is still not expressed in terms of the integration variables. In order to do this, the properties of spherical harmonics and couplings between angular momenta must be used~\cite{QTAM} in such a way that
\begin{align}
&
\left\{ \left[\mathcal{Y}_{L_{\lambda_A}}(\Vec{p}_{\lambda_A}) \mathcal{Y}_{L_{\rho}}(\Vec{p}_{\rho})\right]_{L_A} \mathcal{Y}_1 \left( \Vec{p} \right) \right\}_L = \sum_{l_1,l_2,l_3,l_4,l_5} B^{l_4}_{l_1,l_2} B^{l_5}_{L_{\lambda_A} - l_1,1 - l_2} C^{L_{\lambda_A}}_{l_1} C^{1}_{l_2} \nonumber\\
&
\times \Pi_{L_A,L_{\lambda_A},l_3,l_4,l_5,1} (1-x)^{l_1} x^{l_2} (-1)^{L + L_A + L_{\lambda_A} - l_1 + 1} k^{l_1+l_2-l_4} q^{L_{\lambda_A} - l_1 - l_2 - l_5 + 1} \nonumber\\
&
\times \begin{Bmatrix}
L_{\lambda_A} && L_\rho && L_A\\
L && 1 && l_3
\end{Bmatrix} \begin{Bmatrix}
l_1 && L_{\lambda_A} - l_1 && L_{\lambda_A}\\
l_2 && 1 - l_2 && 1\\
l_4 && l_5 && l_3
\end{Bmatrix} \left\{ \mathcal{Y}_{L_{\rho}}(\Vec{p}_{\rho}) \left[ \mathcal{Y}_{l_4}(\Vec{k}) \mathcal{Y}_{l_5}(\Vec{q}\,) \right]_{l_3} \right\}_L,
\end{align}
where we have defined the following coefficients
\begin{equation}
\begin{matrix}
B_{a,b}^c = (-1)^c \sqrt{\frac{(2a + 1)(2b + 1)}{4\pi}} \begin{pmatrix}
a && b && c\\
0 && 0 && 0
\end{pmatrix}\,, & \quad\quad & C^a_b = \sqrt{\frac{4\pi (2a+1)!}{(2b+1)!(2(a-b)+1)!}} \,.
\end{matrix}
\end{equation}
The remaining term becomes
\begin{align}
&
\left\{ \left[ \left[\mathcal{Y}_{L_{\lambda_B}}(\Vec{p}_{\lambda_B}) \mathcal{Y}_{L_{\rho}}(\Vec{p}_{\rho})\right]_{L_B} \mathcal{Y}_{L_C}(\Vec{p}_C) \right]_{L_{BC}} \mathcal{Y}_l(\Vec{k}) \right\}_L = \nonumber\\ 
&
= \sum_{l_6,l_7,l_8,l_9,l_{10},l_{11},l_{12}} B_{l_6,l_7}^{l_9} B_{L_{\lambda_B} - l_6,L_C - l_7}^{l_{10}} B_{l_9,l}^{l_{12}} C^{L_{\lambda_B}}_{l_6} C^{L_C}_{l_7} \nonumber\\
&
\times \Pi_{L_{BC},L_B,L_C,L_{\lambda_B},l_8,l_8,l_9,l_{10},l_{11},l_{12}}\left(\frac{\omega_\mu}{\omega_{\alpha\beta\mu}} - x\right)^{l_6}  \left(\frac{\omega_\mu}{\omega_{\eta\mu}} - x\right)^{l_7} \nonumber\\
&
\times (-1)^{L_{BC} + L_B + L_{\lambda_B} + L_\rho + L - l_6 - l_7 + l_{10} + l_{12}} k^{l + l_6 + l_7 - l_{12}} q^{L_{\lambda_B} + L_C - l_6 - l_7 - l_{10}} \nonumber\\
&
\times \begin{Bmatrix}
            L_{\lambda_B} && L_\rho && L_B\\
            L_{BC} && L_C && l_8
        \end{Bmatrix} \begin{Bmatrix}
            L_\rho && l_8 && L_{BC}\\
            l && L && l_{11}
        \end{Bmatrix} 
\begin{Bmatrix}
            l_{10} && l_9 && l_8\\
            l && l_{11} && l_{12}
        \end{Bmatrix} \begin{Bmatrix}
            l_6 && L_{\lambda_B} - l_6 && L_{\lambda_B}\\
            l_7 && L_C - l_7 && L_C\\
            l_9 && l_{10} && l_8
        \end{Bmatrix} \nonumber\\
&
\times \left\{ \mathcal{Y}_{L_{\rho}}(\Vec{p}_{\rho}) \left[\mathcal{Y}_{l_{12}}(\Vec{k}) \mathcal{Y}_{l_{10}}(\Vec{q}\,) \right]_{l_{11}} \right\}_L \,.
\end{align}

We arrive then to the expression of the linear momentum contribution
\begin{align}
&
\mathcal{E}(A_{ik} \rightarrow B_{jk}C_{l'}) = \sum_{ijkl'} d_i^{\lambda_A} d_j^{\lambda_B} (d_k^\rho)^2 d_{l'}^C \int d^3q d^3k d^3p_{\rho} \frac{\delta(k-k_0)}{k^{l+1}} \exp\left( - Bq^2 - Dk^2  -\frac{\rho_k}{2} p_\rho^2 \right) \nonumber\\
&
\times \sum_{l_1,l_2,...,l_{11},l_{12}} B^{l_4}_{l_1,l_2} B^{l_5}_{L_{\lambda_A} - l_1,1 - l_2} B_{l_6,l_7}^{l_9} B_{L_{\lambda_B} - l_6,L_C - l_7}^{l_{10}} B_{l_9,l}^{l_{12}} C^{L_{\lambda_A}}_{l_1} C^{1}_{l_2} C^{L_{\lambda_B}}_{l_6} C^{L_C}_{l_7} \nonumber\\
&
\times \Pi_{L_{BC},L_A,L_B,L_C,L_{\lambda_A},L_{\lambda_B},l_3,l_4,l_5,l_8,l_8,l_9,l_{10},l_{11},l_{12},1} \, (1-x)^{l_1} x^{l_2} \left(\frac{\omega_\mu}{\omega_{\alpha\beta\mu}} - x\right)^{l_6}  \left(\frac{\omega_\mu}{\omega_{\eta\mu}} - x\right)^{l_7} \nonumber\\
&
\times (-1)^{L_{BC} + L_A + L_B + L_{\lambda_A} + L_{\lambda_B} + L_\rho - l_1 - l_6 - l_7 + l_{10} + l_{12} + 1} \nonumber\\ 
&
\times k^{l+l_1+l_2-l_4 + l_6 + l_7 - l_{12}} q^{L_C + L_{\lambda_A} + L_{\lambda_B} - l_1 - l_2 - l_5 - l_6 - l_7 - l_{10} + 1} \nonumber\\
&
\times \begin{Bmatrix}
            L_{\lambda_A} && L_\rho && L_A\\
            L && 1 && l_3
        \end{Bmatrix} \begin{Bmatrix}
            L_{\lambda_B} && L_\rho && L_B\\
            L_{BC} && L_C && l_8
        \end{Bmatrix} \begin{Bmatrix}
            L_\rho && l_8 && L_{BC}\\
            l && L && l_{11}
        \end{Bmatrix} \begin{Bmatrix}
            l_{10} && l_9 && l_8\\
            l && l_{11} && l_{12}
        \end{Bmatrix} \nonumber \\
&
\times \begin{Bmatrix}
            l_1 && L_{\lambda_A} - l_1 && L_{\lambda_A}\\
            l_2 && 1 - l_2 && 1\\
            l_4 && l_5 && l_3
        \end{Bmatrix} \begin{Bmatrix}
            l_6 && L_{\lambda_B} - l_6 && L_{\lambda_B}\\
            l_7 && L_C - l_7 && L_C\\
            l_9 && l_{10} && l_8
        \end{Bmatrix} \nonumber\\
&
\times \left\{ \mathcal{Y}_{L_{\rho}}(\Vec{p}_{\rho}) \left[\mathcal{Y}_{l_{12}}(\Vec{k}) \mathcal{Y}_{l_{10}}(\Vec{q}\,) \right]_{l_{11}} \right\}_L^* \left\{ \mathcal{Y}_{L_{\rho}}(\Vec{p}_{\rho}) \left[ \mathcal{Y}_{l_4}(\Vec{k}) \mathcal{Y}_{l_5}(\Vec{q}\,) \right]_{l_3} \right\}_L \,,
\end{align}
in which the angular integrals can be solved using the orthogonality of spherical harmonics as follows~\cite{QTAM}:
\begin{align}
&
\int d^3q\, d^3k\, d^3p_{\rho} \left\{ \mathcal{Y}_{L_{\rho}}(\Vec{p}_{\rho}) \left[\mathcal{Y}_{l_{12}}(\Vec{k}) \mathcal{Y}_{l_{10}}(\Vec{q}\,) \right]_{l_{11}} \right\}_L^* \left\{ \mathcal{Y}_{L_{\rho}}(\Vec{p}_{\rho}) \left[ \mathcal{Y}_{l_4}(\Vec{k}) \mathcal{Y}_{l_5}(\Vec{q}\,) \right]_{l_3} \right\}_L = \nonumber\\
&
= \delta_{L_\rho,L_\rho} \delta_{l_{12},l_4} \delta_{l_{10},l_5} \delta_{l_{11},l_3} \delta_{L,L} \int dq\, dk\, dp_{\rho}\, q^{2 + l_5 + l_{10}} k^{2 + l_4 + l_{12}} p_{\rho}^{2 + 2L_\rho}.
\end{align}
Additionally, the radial integrals can be simplified using the Gamma function when the exponential term is taken into account. The integral over $q$ becomes
\begin{align}
\int_0^\infty dq\, e^{-B q^2} &q^{L_C + L_{\lambda_A} + L_{\lambda_B} - l_1 - l_2 - l_5 - l_6 - l_7 + l_5 + 3} = \nonumber\\
&
= \frac{1}{2} B^{-\frac{1}{2}(L_C + L_{\lambda_A} + L_{\lambda_B} - l_1 - l_2 - l_6 - l_7 + 4)} \nonumber\\
&
\times \Gamma\left( \frac{1}{2}(L_C + L_{\lambda_A} + L_{\lambda_B} - l_1 - l_2 - l_6 - l_7 + 4)\right) \,.
\end{align}
The same can be made for the integral over $p_\rho$,
\begin{equation}
\int dp_{\rho}\, e^{-\frac{\rho_k}{2} p_\rho^2} p_{\rho}^{2 + 2L_\rho} = \frac{1}{2} \left( \frac{\rho_k}{2} \right)^{-\frac{1}{2}(2L_\rho + 3)} \Gamma\left( \frac{1}{2}(2L_\rho + 3)\right) \,,
\end{equation}
and the integral over $k$ can be simplified using the delta-function $\delta(k-k_0)$. 

Threfore, the lineal momentum contribution has the final expression
\begin{align}
&
\mathcal{E}(A_{ik} \rightarrow B_{jk}C_{l'}) = \sum_{ijkl'} d_i^{\lambda_A} d_j^{\lambda_B} (d_k^\rho)^2 d_{l'}^C  \exp\left( - Dk_0^2 \right) \nonumber\\
&
\times \sum_{l_1,l_2,...,l_8,l_9} B^{l_4}_{l_1,l_2} B^{l_5}_{L_{\lambda_A} - l_1,1 - l_2} B_{l_6,l_7}^{l_9} B_{L_{\lambda_B} - l_6,L_C - l_7}^{l_5} B_{l_9,l}^{l_4} C^{L_{\lambda_A}}_{l_1} C^{1}_{l_2} C^{L_{\lambda_B}}_{l_6} C^{L_C}_{l_7} \nonumber\\
&
\times \Pi_{L_{BC},L_A,L_B,L_C,L_{\lambda_A},L_{\lambda_B},l_3,l_3,l_4,l_4,l_5,l_5,l_8,l_8,l_9,1} (1-x)^{l_1} x^{l_2} \left(\frac{\omega_\mu}{\omega_{\alpha\beta\mu}} - x\right)^{l_6}  \left(\frac{\omega_\mu}{\omega_{\eta\mu}} - x\right)^{l_7} \nonumber\\
&
\times (-1)^{L_{BC} + L_A + L_B + L_{\lambda_A} + L_{\lambda_B} + L_\rho - l_1 - l_6 - l_7 + l_5 + l_4 + 1} (2)^{L_\rho - \frac{1}{2}} \nonumber\\
&
\times k_0^{l_1+l_2 + l_6 + l_7 + 1} B^{-\frac{1}{2}(L_C + L_{\lambda_A} + L_{\lambda_B} - l_1 - l_2 - l_6 - l_7 + 4)} \rho_k^{-\frac{1}{2}(2L_\rho + 3)} \nonumber\\
&
\times \Gamma\left( \frac{1}{2}(L_C + L_{\lambda_A} + L_{\lambda_B} - l_1 - l_2 - l_6 - l_7 + 4)\right) \Gamma\left( \frac{1}{2}(2L_\rho + 3)\right) \nonumber\\
&
\times \begin{Bmatrix}
            L_{\lambda_A} && L_\rho && L_A\\
            L && 1 && l_3
        \end{Bmatrix} \begin{Bmatrix}
            L_{\lambda_B} && L_\rho && L_B\\
            L_{BC} && L_C && l_8
        \end{Bmatrix} \begin{Bmatrix}
            L_\rho && l_8 && L_{BC}\\
            l && L && l_3
        \end{Bmatrix} \begin{Bmatrix}
            l_5 && l_9 && l_8\\
            l && l_3 && l_4
        \end{Bmatrix} \nonumber \\
&
\times \begin{Bmatrix}
            l_1 && L_{\lambda_A} - l_1 && L_{\lambda_A}\\
            l_2 && 1 - l_2 && 1\\
            l_4 && l_5 && l_3
        \end{Bmatrix} \begin{Bmatrix}
            l_6 && L_{\lambda_B} - l_6 && L_{\lambda_B}\\
            l_7 && L_C - l_7 && L_C\\
            l_9 && l_5 && l_8
        \end{Bmatrix} \,,
        \label{SSFinal}
\end{align}
where the limits of the sums can be obtained using triangular conditions of the Wigner symbols.

\subsection{The color contribution}

Concerning the color matrix element, this can be written as
\begin{equation}
\mathcal{I}_{Color} = \langle C_B C_C | C_A \rangle = \langle C(\epsilon\sigma\zeta) C(\delta\tau) | C(\alpha\beta\eta) C(\mu\nu) \rangle \,,
\label{eq:color}
\end{equation}
where the color function of the pair created is added. In order to calculate this contribution, the color function of the baryons and mesons must be known:
\begin{align}
C(\alpha\beta\eta) &= \frac{1}{\sqrt{6}} \sum_{\alpha\beta\eta} \varepsilon_{\alpha\beta\eta}  = \frac{1}{\sqrt{6}} (r_\alpha g_\beta b_\eta - r_\alpha b_\beta g_\eta + g_\alpha b_\beta r_\eta - g_\alpha r_\beta b_\eta + b_\alpha r_\beta g_\eta - b_\alpha g_\beta r_\eta) \,, \\
C(\delta\tau) &= \frac{1}{\sqrt{3}} \sum_{\delta\tau} \delta_{\delta\tau} = \frac{1}{\sqrt{3}}(r_\delta \overline{r}_\tau + g_\delta \overline{g}_\tau + b_\delta \overline{b}_\tau) \,.
\end{align}
Since mesons are made of a quark and antiquark a Kronecker delta is enough to describe the color function. On the other hand, the baryon must have an antisymmetric color wave function, this make the Levi-Civita symbol sufficient to describe the color of the system.

The color wave functions are replaced in Eq.~\eqref{eq:color}, taking into account the particle equivalences of Eq.~\eqref{Equiva}, the color contribution becomes
\begin{align}
\mathcal{I}_{Color} &= \frac{1}{18} \sum_{\alpha\beta\eta} \sum_{\mu\nu} \sum_{\epsilon\sigma\zeta} \sum_{\delta\tau} \varepsilon_{\alpha\beta\eta} \delta_{\mu\nu} \varepsilon_{\epsilon\sigma\zeta} \delta_{\delta\tau} \nonumber\\
&
= \frac{1}{18} \sum_{\alpha\beta\eta\mu} \sum_{\epsilon\sigma\zeta\delta} \varepsilon_{\alpha\beta\eta} \varepsilon_{\epsilon\sigma\zeta} \delta_{\delta\mu} \delta_{\epsilon\mu} \delta_{\zeta\alpha} \delta_{\sigma\beta} \delta_{\delta\eta} \nonumber\\
&
= \frac{1}{18} \sum_{\alpha\beta\eta} \sum_{\delta} \varepsilon_{\alpha\beta\eta} \varepsilon_{\delta\beta\alpha} \delta_{\delta\eta} \nonumber\\
&
= \frac{1}{18} \sum_{\alpha\beta\eta} \varepsilon_{\alpha\beta\eta} \varepsilon_{\eta\beta\alpha} \,.
\end{align}
Now, the product of Levi-Civita tensors can be simplified as
\begin{equation}
\sum_{\alpha\beta\eta} \varepsilon_{\alpha\beta\eta} \varepsilon_{\eta\beta\alpha} = -6 \,,
\end{equation}
arriving at 
\begin{equation}
\mathcal{I}_{Color} = -\frac{1}{3} \,.
\end{equation}
This term cancels with the $3$ put by hand in the transition operator, Eq.~\eqref{1Top}.

\subsection{The flavor contribution}

The flavor matrix element may be written as
\begin{equation}
\mathcal{I}_{Sabor} = \langle [(t_\mu t_\beta t_\alpha)I_B(t_\nu t_\eta)I_C]I_A | [(t_\alpha t_\beta t_\eta)I_A (t_\mu t_\nu) 0]I_A \rangle \,.
\end{equation}
In order to simplify this expression, the flavor of the non-interacting quarks ($\alpha$ and $\beta$) inside baryons is defined as already done for spin,
\begin{equation}
t_\rho = t_\alpha \otimes t_\beta.
\end{equation}
Thus, the final expression to be calculated is
\begin{equation}
\mathcal{I}_{Sabor} = (-1)^{t_\alpha + t_\beta + t_\mu - I_B} (-1)^{t_\nu + t_\eta - I_C} \langle [(t_\rho t_\mu)I_B(t_\eta t_\nu)I_C]I_A | [(t_\rho t_\eta)I_A (t_\mu t_\nu) 0]I_A \rangle \,.
\end{equation}
This can be re-written using a $9J$-symbol as
\begin{equation}
\mathcal{I}_{Sabor} = (-1)^{t_\alpha + t_\beta + t_\mu - I_B} (-1)^{t_\nu + t_\eta - I_C} \Pi_{I_B,I_C,I_A,0} \begin{Bmatrix}
t_{\rho} && t_\mu && I_B\\
t_\eta && t_\nu && I_C\\
I_A && 0 && I_A
\end{Bmatrix} \,.
\end{equation}
Since the $9J$-symbol has a zero in one of its components, it can be simplified into a $6J$-symbol,
\begin{equation}
\mathcal{I}_{Sabor} =(-1)^{t_\alpha + t_\beta + t_\mu + t_\eta + t_\rho + I_A - I_B} \frac{\Pi_{I_B,I_C}}{\Pi_{t_\mu}} \begin{Bmatrix}
t_\eta && I_C && t_\mu\\
I_B && t_{\rho_A} && I_A
\end{Bmatrix},
\end{equation}
where the equivalence of $t_\mu=t_\nu$ is used.


\section{Results}
\label{sec:results}

Once we have shown a detailed derivation of the analytical expression for the decay width, and transition matrix, of a baryon decaying strongly into a meson plus another baryon, it is time to provide an example of calculation in order to check the correctness of it. Besides, one of our long-term goals is to provide a unified picture of mesons and baryons decaying strongly, from our previous analysis in the meson sector~\cite{Segovia:2012cd}. 

The most convenient example for our test is the $\Delta(1232)$ baryon decaying strongly into a pion ($\pi(140)$) plus a nucleon ($N(940)$). This is because (i) all hadrons involved in the reaction are very well established in the Particle Listings of Particle Data Group (PDG)~\cite{Workman:2022ynf}; (ii) the two hadrons in the final state are stable avoiding additional complications in the computation related with taking into account decay widths of the products; (iii) all constituent quarks inside hadrons are either $u$- or $d$-quarks and, since isospin symmetry is well fulfilled in QCD, one can assume all as equivalent quarks; (iv) the branching fraction of the $\Delta(1232)\to \pi N$ strong decay channel is $99.4\%$ which constitutes almost the total decay width; and (v) the total decay width is relatively well measured experimentally, with a value between $114\,\text{MeV}$ and $120\,\text{MeV}$~\cite{Workman:2022ynf}.\footnote{Note the use of the so-called natural units, $\hbar=c=1$.}

As we have already mentioned, the $\Delta$-baryon is made of up (u) and down (d) quarks in different combinations, this make $4$ different species with different charges and decay channels. This species an their properties can be resumed in the following table,
\begin{table}[H]
    \centering
    \begin{tabular}{c|c|c|c}
         \textbf{ Baryon } & \textbf{ Quarks } & \textbf{ Charge (in units of e) }  & \textbf{ Decay Channels }\\
         $\Delta^{++}$ & uuu & +2 & p$^+$ + $\pi^+$\\
         $\Delta^{+}$ & uud & +1 & n$^0$ + $\pi^+$ or p$^+$ + $\pi^0$ \\
         $\Delta^{0}$ & udd & 0 & n$^0$ + $\pi^+$ or p$^+$ + $\pi^-$ \\
         $\Delta^{-}$ & ddd & -1 & n$^0$ + $\pi^-$ \\
    \end{tabular}
    \label{Especie}
\end{table}

In order to calculate the decay width, the properties of the initial and final hadrons must be fixed. To simplify, and without loss of generality, the studied decaying baryon is $\Delta(1232)^-$. Therefore, the properties of all the hadrons involved in the reaction are listed as (a constituent quark model description of hadrons is assumend):
\begin{table}[H]
    \centering
    \begin{tabular}{c | c  | c  | c  | c | c }
         \textbf{ Hadron } & $ \textbf{L} $ & $\textbf{S}$ & $\textbf{J}$ & \textbf{ Mass (MeV) } & \textbf{ Radius (fm) } \\
         $\Delta$ & $0$ & $3/2$ & $3/2$ & $1232$ & $1.03$ \\
         n & $0$ & $1/2$ & $1/2$ & $940$ & $0.84$ \\
         $\pi^{+/-}$ & $0$ & $0$ & $0$ & $140$ & $0.5$\\
    \end{tabular}
    \label{tab:Porp}
\end{table}

Using the experimental masses, the transferred momentum of the reaction can be calculated:
\begin{equation}
k_0 = \frac{\sqrt{\left(m_A^2 - (m_B - m_C)^2\right)\left(m_A^2 - (m_B + m_C)^2\right)}}{2m_A} = 226\,\text{MeV} \,.
\end{equation}

Having fixed the target reaction to be studied, let us now disentangle some relevant couplings needed to provide a final numerical result. For example, the coupling of angular momenta in the final state can be deduced as follows
\begin{equation}
\begin{matrix}
J_{BC} = J_B \otimes J_C = \frac{1}{2}& \, & \text{and} & \, & L_{BC} = L_B \otimes L_C = 0 \,.
\end{matrix}
\end{equation}
For the other values is necessary a little more of work. For instance, the baryons have two internal momenta that comes form the $(\rho\lambda)$-Jacobi coordinates, they should fulfill
\begin{align}
L_{\lambda_A} \otimes L_\rho = 0 \,, \nonumber\\
L_{\lambda_B} \otimes L_\rho = 0 \,,
\end{align}
which dictate that
\begin{equation}
L_{\lambda_A} = L_{\lambda_B} = L_\rho \,,
\end{equation}
and we assume that all are zero. Besides, the final hadrons have a relative angular momentum, $l$, which must be taken into account to assure the conservation of angular momentum,
\begin{equation}
\Vec{J}_A = \Vec{J}_{BC} + \Vec{l} \quad
\Rightarrow \quad \frac{3}{2} = \left|\frac{1}{2} \otimes l\right|,
\end{equation}
having two possible values,
\begin{equation}
l = 1 \text{ or } l = 2 \,.
\end{equation}
During the calculation the Wigner symbols eliminate any of the values that do not comply with the triangular conditions.

Concerning spin coupling, the quarks are fermions with spin $1/2$ and $s_\rho$ must be maintained since it is the total spin of the two spectator quarks. Therefore, the next decomposition is made,
\begin{align}
S_A &= \frac{3}{2} = \left( \frac{1}{2} \otimes \frac{1}{2}\right)_1 \otimes \frac{1}{2} = \left( s_\alpha \otimes s_\beta \right)_{s_\rho} \otimes s_\eta \,, \nonumber\\
S_B &= \frac{1}{2} = \left( \frac{1}{2} \otimes \frac{1}{2}\right)_1 \otimes \frac{1}{2} = \left( s_\alpha \otimes s_\beta \right)_{s_\rho} \otimes s_\mu \,, \nonumber\\
S_C &= 0 = \frac{1}{2} \otimes \frac{1}{2}  = s_\eta \otimes s_\nu \,.
\end{align}
Note that the spin of the spectator quarks must be equal to $1$ in order to provide correctly the $\Delta$'s quantum numbers and thus this requires the same value of $s_\rho$ in the nucleon. Similar reasoning is made for the isospin couplings:
\begin{align}
I_A &= \frac{3}{2} = \left( \frac{1}{2} \otimes \frac{1}{2}\right)_{1} \otimes \frac{1}{2} = \left( t_\alpha \otimes t_\beta \right)_{t_\rho} \otimes t_\eta \,, \nonumber\\
I_B &= \frac{1}{2} = \left( \frac{1}{2} \otimes \frac{1}{2}\right)_{1} \otimes \frac{1}{2} = \left( t_\alpha \otimes t_\beta \right)_{t_\rho} \otimes t_\mu \,, \nonumber\\
I_C &= 1 = \frac{1}{2} \otimes \frac{1}{2}  = t_\eta \otimes t_\nu \,.
\end{align}

In order to estimate the coefficients that appear in the Gaussian expansion of the hadron's wave functions, some analysis is needed. The Gaussian form for the meson can be assumed to be,
\begin{equation}
f(\Vec{p}_C) = d^C e^{-\frac{C}{2}p^2_C} \,,
\end{equation}
where only one term of the sum is used in order to simplify the calculation. The form of the Gaussian function gives the next relation between the variance and the coefficient,
\begin{equation}
\sigma^2 = \frac{1}{C} \,.
\end{equation}
Then, the following approximation can be done for the hadron's radius,
\begin{equation}
\langle r^2 \rangle \approx \frac{(\hbar c)^2}{\sigma^2},
\end{equation}
where $\hbar c = 0.197327\,\text{GeV fm}$ is added to have the correct units. This is the relationship that we are going to use between hadron's coefficient and its size.

The Gaussian expansion for a baryon is different since it has two components. Following Ref.~\cite{Straub:1988gj} the next function is used,
\begin{equation}
f(\Vec{p}_\lambda,\Vec{p}_\rho) = \left[ \frac{2b^2}{\pi} \right]^{\frac{3}{4}} e^{-b^2\Vec{p\,}_\rho^2} \left[ \frac{2\alpha b^2}{\pi} \right]^{\frac{3}{4}} e^{-\alpha b^2\Vec{p\,}_\lambda^2} \,,
\label{eq:fbaryon}
\end{equation}
where $\alpha$ depends on quark masses as
\begin{equation}
\alpha = \frac{m_1m_2(m_1 + m_2 + m_3)}{m_3(m_1 + m_2)^2} \,,
\end{equation}
which is equal to $3/4$ in our case. Equation~\eqref{eq:fbaryon} shows us two important aspects: (i) the computation of the baryon radius in terms of its Gaussian's standard deviation and (ii) the relation
\begin{equation}
d^C = \left[\frac{2C}{\pi}\right]^{3/4} \,.
\end{equation}

Now, we are in the position of computing the coefficients necessary for the transition matrix (see Eqs.~\eqref{eq:C1}-\eqref{eq:C4}),
\begin{align}
A & =  \frac{9\langle r_A \rangle^2 + 3\langle r_B \rangle^2 + 2\langle r_C \rangle^2}{4(\hbar c)^2} \,,\\
B & = \frac{9\langle r_A \rangle^2 + 9\langle r_B \rangle^2 + 4\langle r_C \rangle^2}{8(\hbar c)^2} \,,\\
x & = \frac{9\langle r_A \rangle^2 + 3\langle r_B \rangle^2 + 2\langle r_C \rangle^2}{9\langle r_A \rangle^2 + 9\langle r_B \rangle^2 + 4\langle r_C \rangle^2} \,,\\
D &= \frac{1}{2}\left( \frac{9\langle r_A \rangle^2}{4(\hbar c)^2}\left(1-x \right)^2 + \frac{9\langle r_B \rangle^2}{4(\hbar c)^2} \left(\frac{1}{3} - x\right)^2 + \frac{\langle r_C \rangle^2}{(\hbar c)^2}\left(\frac{1}{2} - x\right)^2 \right) \,,
\end{align}
and the product of the amplitudes,
\begin{equation}
d_i^{\lambda_A} d_i^{\lambda_B}(d_k^\rho)^2 d_l^C = \rho_k^{\frac{3}{2}} \left[\left(\frac{2}{\pi}\right)^{5} \left(\frac{81\langle r_A \rangle^2 \langle r_B \rangle^2 \langle r_C \rangle^2}{16(\hbar c)^6} \right) \right]^{\frac{3}{4}} \,.
\end{equation}
Note that $\rho_k^{\frac{3}{2}}$ cancels with the $\rho_k^{-\frac{3}{2}}$ that appears in the spin-space component, Eq.~\eqref{SSFinal}.

The final constant that needs to be known is only-free paramter of the ${}^3P_0$ decay model, $\gamma'$, that characterizes the strength of the quark-antiquark pair creation from the vacuum. Following Ref.~\cite{Segovia:2012cd}, the next relation can be used to calculate this constant,
\begin{equation}
\gamma' = \sqrt{2^5 \pi} \,\cdot  \frac{\gamma_0}{\log\left(\frac{\mu}{\mu_\gamma}\right)},
\end{equation}
where $\gamma_0=0.81\pm0.02$ and $\mu_\gamma=49.84\pm2.58\,\text{MeV}$ are constants fitted to the total strong decay widths of mesons, and $\mu$ represents the quark sector to which the decaying hadron belongs. For the example at hand, the $\Delta$-baryon belongs to the light quark sector; therefore, we have $\mu=156.5\,\text{MeV}$ and
\begin{equation}
\gamma'= \sqrt{16\pi} \approx 7.09 \,.
\end{equation}

All together provide the following value for the decay width
\begin{equation}
\Gamma({\Delta(1232)^- \to n(940)+\pi(140)^-}) = 113.32\,\text{MeV} \,,
\end{equation}
which is just at, or below, the minimum given by the PDG~\cite{Workman:2022ynf}, \emph{viz.} $\Gamma_{\Delta(1232)}=(114-120)\,\text{MeV}$ and so $\Gamma_{\Delta(1232)\to N\pi}=(113-119)\,\text{MeV}$. This result is quite remarkable since, in some sense, it is a free-parameter prediction of the ${}^3P_0$ decay model using just the experimental values of the hadron's radii reported in Ref.~\cite{Workman:2022ynf} and the scale-dependent strength determined in Ref.~\cite{Segovia:2012cd}.

To contrast this result, in the literature, the $\gamma'$ value usually used is the one obtained in the fitting made by Blundell~\cite{Blundell:1995ev}. Since this value is fixed for meson decays, it must be divided by $\sqrt{3}$ to extrapolate it for baryon decays. Therefore, the next value of $\gamma'$ can be used, 
\begin{equation}
\gamma' = \frac{13.4}{\sqrt{3}},
\end{equation}
having a decay width of
\begin{equation}
\Gamma({\Delta(1232)^- \to n(940)+\pi(140)^-}) = 134.95\,\text{MeV} \,,
\end{equation}
which is slightly higher than the experimental interval but relatively correct~\cite{Workman:2022ynf}, confirming that the analytical development of the ${}^3P_0$ model for baryon decays seems correct and the radii of the involved hadrons also well estimated.


\section{Summary}
\label{sec:summary}

This work have shown how to obtain in detail the transition matrix element for a baryon decaying strongly into a meson and another baryon through the well-known $^3P_0$ quark-antiquark pair creation model. 

Since one of our long-term goals is to provide a unified picture of mesons and baryons from our chiral quark model, one important feature is to describe under the same umbrella meson and baryons strong decays. The meson's study was conducted in Ref.~\cite{Segovia:2012cd} and we have wanted to extend the same formalism to the baryon sector, focusing on the $\Delta(1232)\to \pi N$ strong decay width because all hadrons involved in the reaction are very well established, the two hadrons in the final state are stable avoiding further analysis, all quarks are light and equivalent, and the decay width of the process is relatively well measured. 

Taking advantage of a Gaussian expansion method for the hadron's radial wave functions, the expression of the invariant matrix element can be simplified into a sum of multiple terms composed basically on some numerical values, wavefunction coefficients and Wigner symbols. Those wavefunction coefficients can be determined from the mean-square radii of involved hadrons, and we have used their experimental measures in such a way that the only one free parameter is the strength of the quark-antiquark pair creation from the vacuum. This has been taken from our previous study of strong decay widths in the meson sector~\cite{Segovia:2012cd} and we have obtained a quite compatible result with experiment for the calculated $\Delta(1232)\to \pi N$ decay width.

Aware of our weaknesses, we are working on applying the model developed herein to more baryon strong decays, but to do so we must first develop a numerical method that solves the bound state problem for baryons. This is where we are putting our efforts right now.


\begin{acknowledgments}
Work partially financed by 
the Escuela Polit\'ecnica Nacional under projects PIS-22-04 and PIM 19-01; 
EU Horizon 2020 research and innovation program, STRONG-2020 project, under grant agreement no. 824093;
Ministerio Espa\~nol de Ciencia e Innovaci\'on under grant nos. PID2022-141910NB-I00 and PID2022-140440NB-C22;
Junta de Andaluc\'ia under contract no. PAIDI FQM-370.
\end{acknowledgments}


\bibliography{print3P0Baryons}

\begin{thebibliography}{43}%
\makeatletter
\providecommand \@ifxundefined [1]{%
 \@ifx{#1\undefined}
}%
\providecommand \@ifnum [1]{%
 \ifnum #1\expandafter \@firstoftwo
 \else \expandafter \@secondoftwo
 \fi
}%
\providecommand \@ifx [1]{%
 \ifx #1\expandafter \@firstoftwo
 \else \expandafter \@secondoftwo
 \fi
}%
\providecommand \natexlab [1]{#1}%
\providecommand \enquote  [1]{``#1''}%
\providecommand \bibnamefont  [1]{#1}%
\providecommand \bibfnamefont [1]{#1}%
\providecommand \citenamefont [1]{#1}%
\providecommand \href@noop [0]{\@secondoftwo}%
\providecommand \href [0]{\begingroup \@sanitize@url \@href}%
\providecommand \@href[1]{\@@startlink{#1}\@@href}%
\providecommand \@@href[1]{\endgroup#1\@@endlink}%
\providecommand \@sanitize@url [0]{\catcode `\\12\catcode `\$12\catcode
  `\&12\catcode `\#12\catcode `\^12\catcode `\_12\catcode `\%12\relax}%
\providecommand \@@startlink[1]{}%
\providecommand \@@endlink[0]{}%
\providecommand \url  [0]{\begingroup\@sanitize@url \@url }%
\providecommand \@url [1]{\endgroup\@href {#1}{\urlprefix }}%
\providecommand \urlprefix  [0]{URL }%
\providecommand \Eprint [0]{\href }%
\providecommand \doibase [0]{https://doi.org/}%
\providecommand \selectlanguage [0]{\@gobble}%
\providecommand \bibinfo  [0]{\@secondoftwo}%
\providecommand \bibfield  [0]{\@secondoftwo}%
\providecommand \translation [1]{[#1]}%
\providecommand \BibitemOpen [0]{}%
\providecommand \bibitemStop [0]{}%
\providecommand \bibitemNoStop [0]{.\EOS\space}%
\providecommand \EOS [0]{\spacefactor3000\relax}%
\providecommand \BibitemShut  [1]{\csname bibitem#1\endcsname}%
\let\auto@bib@innerbib\@empty
\bibitem [{\citenamefont {Segovia}\ \emph
  {et~al.}(2012{\natexlab{a}})\citenamefont {Segovia}, \citenamefont {Entem},\
  and\ \citenamefont {Fern\'andez}}]{Segovia:2012cd}%
  \BibitemOpen
  \bibfield  {author} {\bibinfo {author} {\bibfnamefont {J.}~\bibnamefont
  {Segovia}}, \bibinfo {author} {\bibfnamefont {D.~R.}\ \bibnamefont {Entem}},\
  and\ \bibinfo {author} {\bibfnamefont {F.}~\bibnamefont {Fern\'andez}},\
  }\bibfield  {title} {\bibinfo {title} {{Scaling of the $^3P_0$ Strength in
  Heavy Meson Strong Decays}},\ }\href
  {https://doi.org/10.1016/j.physletb.2012.08.005} {\bibfield  {journal}
  {\bibinfo  {journal} {Phys. Lett. B}\ }\textbf {\bibinfo {volume} {715}},\
  \bibinfo {pages} {322} (\bibinfo {year} {2012}{\natexlab{a}})},\ \Eprint
  {https://arxiv.org/abs/1205.2215} {arXiv:1205.2215 [hep-ph]} \BibitemShut
  {NoStop}%
\bibitem [{\citenamefont {Workman}\ \emph {et~al.}(2022)\citenamefont {Workman}
  \emph {et~al.}}]{ParticleDataGroup:2022pth}%
  \BibitemOpen
  \bibfield  {author} {\bibinfo {author} {\bibfnamefont {R.~L.}\ \bibnamefont
  {Workman}} \emph {et~al.} (\bibinfo {collaboration} {Particle Data Group}),\
  }\bibfield  {title} {\bibinfo {title} {{Review of Particle Physics}},\ }\href
  {https://doi.org/10.1093/ptep/ptac097} {\bibfield  {journal} {\bibinfo
  {journal} {PTEP}\ }\textbf {\bibinfo {volume} {2022}},\ \bibinfo {pages}
  {083C01} (\bibinfo {year} {2022})}\BibitemShut {NoStop}%
\bibitem [{\citenamefont {Papavassiliou}(2015)}]{Papavassiliou:2015aga}%
  \BibitemOpen
  \bibfield  {author} {\bibinfo {author} {\bibfnamefont {J.}~\bibnamefont
  {Papavassiliou}},\ }\bibfield  {title} {\bibinfo {title} {{Unraveling the
  organization of the QCD tapestry}},\ }\href
  {https://doi.org/10.1088/1742-6596/631/1/012006} {\bibfield  {journal}
  {\bibinfo  {journal} {J. Phys. Conf. Ser.}\ }\textbf {\bibinfo {volume}
  {631}},\ \bibinfo {pages} {012006} (\bibinfo {year} {2015})},\ \Eprint
  {https://arxiv.org/abs/1503.04212} {arXiv:1503.04212 [hep-ph]} \BibitemShut
  {NoStop}%
\bibitem [{\citenamefont {Gell-Mann}(1964)}]{Gell-Mann:1964ewy}%
  \BibitemOpen
  \bibfield  {author} {\bibinfo {author} {\bibfnamefont {M.}~\bibnamefont
  {Gell-Mann}},\ }\bibfield  {title} {\bibinfo {title} {{A Schematic Model of
  Baryons and Mesons}},\ }\href {https://doi.org/10.1016/S0031-9163(64)92001-3}
  {\bibfield  {journal} {\bibinfo  {journal} {Phys. Lett.}\ }\textbf {\bibinfo
  {volume} {8}},\ \bibinfo {pages} {214} (\bibinfo {year} {1964})}\BibitemShut
  {NoStop}%
\bibitem [{\citenamefont {Zweig}(1964{\natexlab{a}})}]{Zweig:1964CERN}%
  \BibitemOpen
  \bibfield  {author} {\bibinfo {author} {\bibfnamefont {G.}~\bibnamefont
  {Zweig}},\ }\bibfield  {title} {\bibinfo {title} {Developments in the quark
  theory of hadrons},\ }\href@noop {} {\bibfield  {journal} {\bibinfo
  {journal} {CERN Report No.8182/TH.401, CERN Report No.8419/TH.412}\ }
  (\bibinfo {year} {1964}{\natexlab{a}})}\BibitemShut {NoStop}%
\bibitem [{\citenamefont {Fern\'andez}\ and\ \citenamefont
  {Segovia}(2021)}]{Fernandez:2021zjq}%
  \BibitemOpen
  \bibfield  {author} {\bibinfo {author} {\bibfnamefont {F.}~\bibnamefont
  {Fern\'andez}}\ and\ \bibinfo {author} {\bibfnamefont {J.}~\bibnamefont
  {Segovia}},\ }\bibfield  {title} {\bibinfo {title} {{Historical Introduction
  to Chiral Quark Models}},\ }\href {https://doi.org/10.3390/sym13020252}
  {\bibfield  {journal} {\bibinfo  {journal} {Symmetry}\ }\textbf {\bibinfo
  {volume} {13}},\ \bibinfo {pages} {252} (\bibinfo {year} {2021})}\BibitemShut
  {NoStop}%
\bibitem [{\citenamefont {Vijande}\ \emph {et~al.}(2005)\citenamefont
  {Vijande}, \citenamefont {Fernandez},\ and\ \citenamefont
  {Valcarce}}]{Vijande:2004he}%
  \BibitemOpen
  \bibfield  {author} {\bibinfo {author} {\bibfnamefont {J.}~\bibnamefont
  {Vijande}}, \bibinfo {author} {\bibfnamefont {F.}~\bibnamefont {Fernandez}},\
  and\ \bibinfo {author} {\bibfnamefont {A.}~\bibnamefont {Valcarce}},\
  }\bibfield  {title} {\bibinfo {title} {{Constituent quark model study of the
  meson spectra}},\ }\href {https://doi.org/10.1088/0954-3899/31/5/017}
  {\bibfield  {journal} {\bibinfo  {journal} {J. Phys. G}\ }\textbf {\bibinfo
  {volume} {31}},\ \bibinfo {pages} {481} (\bibinfo {year} {2005})},\ \Eprint
  {https://arxiv.org/abs/hep-ph/0411299} {arXiv:hep-ph/0411299} \BibitemShut
  {NoStop}%
\bibitem [{\citenamefont {Segovia}\ \emph
  {et~al.}(2013{\natexlab{a}})\citenamefont {Segovia}, \citenamefont {Entem},
  \citenamefont {Fernandez},\ and\ \citenamefont
  {Hernandez}}]{Segovia:2013wma}%
  \BibitemOpen
  \bibfield  {author} {\bibinfo {author} {\bibfnamefont {J.}~\bibnamefont
  {Segovia}}, \bibinfo {author} {\bibfnamefont {D.~R.}\ \bibnamefont {Entem}},
  \bibinfo {author} {\bibfnamefont {F.}~\bibnamefont {Fernandez}},\ and\
  \bibinfo {author} {\bibfnamefont {E.}~\bibnamefont {Hernandez}},\ }\bibfield
  {title} {\bibinfo {title} {{Constituent quark model description of charmonium
  phenomenology}},\ }\href {https://doi.org/10.1142/S0218301313300269}
  {\bibfield  {journal} {\bibinfo  {journal} {Int. J. Mod. Phys. E}\ }\textbf
  {\bibinfo {volume} {22}},\ \bibinfo {pages} {1330026} (\bibinfo {year}
  {2013}{\natexlab{a}})},\ \Eprint {https://arxiv.org/abs/1309.6926}
  {arXiv:1309.6926 [hep-ph]} \BibitemShut {NoStop}%
\bibitem [{\citenamefont {Segovia}\ \emph {et~al.}(2008)\citenamefont
  {Segovia}, \citenamefont {Yasser}, \citenamefont {Entem},\ and\ \citenamefont
  {Fernandez}}]{Segovia:2008zz}%
  \BibitemOpen
  \bibfield  {author} {\bibinfo {author} {\bibfnamefont {J.}~\bibnamefont
  {Segovia}}, \bibinfo {author} {\bibfnamefont {A.~M.}\ \bibnamefont {Yasser}},
  \bibinfo {author} {\bibfnamefont {D.~R.}\ \bibnamefont {Entem}},\ and\
  \bibinfo {author} {\bibfnamefont {F.}~\bibnamefont {Fernandez}},\ }\bibfield
  {title} {\bibinfo {title} {{JPC=1-- hidden charm resonances}},\ }\href
  {https://doi.org/10.1103/PhysRevD.78.114033} {\bibfield  {journal} {\bibinfo
  {journal} {Phys. Rev. D}\ }\textbf {\bibinfo {volume} {78}},\ \bibinfo
  {pages} {114033} (\bibinfo {year} {2008})}\BibitemShut {NoStop}%
\bibitem [{\citenamefont {Segovia}\ \emph
  {et~al.}(2016{\natexlab{a}})\citenamefont {Segovia}, \citenamefont {Ortega},
  \citenamefont {Entem},\ and\ \citenamefont {Fern\'andez}}]{Segovia:2016xqb}%
  \BibitemOpen
  \bibfield  {author} {\bibinfo {author} {\bibfnamefont {J.}~\bibnamefont
  {Segovia}}, \bibinfo {author} {\bibfnamefont {P.~G.}\ \bibnamefont {Ortega}},
  \bibinfo {author} {\bibfnamefont {D.~R.}\ \bibnamefont {Entem}},\ and\
  \bibinfo {author} {\bibfnamefont {F.}~\bibnamefont {Fern\'andez}},\
  }\bibfield  {title} {\bibinfo {title} {{Bottomonium spectrum revisited}},\
  }\href {https://doi.org/10.1103/PhysRevD.93.074027} {\bibfield  {journal}
  {\bibinfo  {journal} {Phys. Rev. D}\ }\textbf {\bibinfo {volume} {93}},\
  \bibinfo {pages} {074027} (\bibinfo {year} {2016}{\natexlab{a}})},\ \Eprint
  {https://arxiv.org/abs/1601.05093} {arXiv:1601.05093 [hep-ph]} \BibitemShut
  {NoStop}%
\bibitem [{\citenamefont {Segovia}\ \emph
  {et~al.}(2015{\natexlab{a}})\citenamefont {Segovia}, \citenamefont {Entem},\
  and\ \citenamefont {Fernandez}}]{Segovia:2015dia}%
  \BibitemOpen
  \bibfield  {author} {\bibinfo {author} {\bibfnamefont {J.}~\bibnamefont
  {Segovia}}, \bibinfo {author} {\bibfnamefont {D.~R.}\ \bibnamefont {Entem}},\
  and\ \bibinfo {author} {\bibfnamefont {F.}~\bibnamefont {Fernandez}},\
  }\bibfield  {title} {\bibinfo {title} {{Charmed-strange Meson Spectrum: Old
  and New Problems}},\ }\href {https://doi.org/10.1103/PhysRevD.91.094020}
  {\bibfield  {journal} {\bibinfo  {journal} {Phys. Rev. D}\ }\textbf {\bibinfo
  {volume} {91}},\ \bibinfo {pages} {094020} (\bibinfo {year}
  {2015}{\natexlab{a}})},\ \Eprint {https://arxiv.org/abs/1502.03827}
  {arXiv:1502.03827 [hep-ph]} \BibitemShut {NoStop}%
\bibitem [{\citenamefont {Segovia}\ \emph {et~al.}(2009)\citenamefont
  {Segovia}, \citenamefont {Yasser}, \citenamefont {Entem},\ and\ \citenamefont
  {Fernandez}}]{Segovia:2009zz}%
  \BibitemOpen
  \bibfield  {author} {\bibinfo {author} {\bibfnamefont {J.}~\bibnamefont
  {Segovia}}, \bibinfo {author} {\bibfnamefont {A.~M.}\ \bibnamefont {Yasser}},
  \bibinfo {author} {\bibfnamefont {D.~R.}\ \bibnamefont {Entem}},\ and\
  \bibinfo {author} {\bibfnamefont {F.}~\bibnamefont {Fernandez}},\ }\bibfield
  {title} {\bibinfo {title} {{Ds-1 (2536) + decays and the properties of P-wave
  charmed strange mesons}},\ }\href
  {https://doi.org/10.1103/PhysRevD.80.054017} {\bibfield  {journal} {\bibinfo
  {journal} {Phys. Rev. D}\ }\textbf {\bibinfo {volume} {80}},\ \bibinfo
  {pages} {054017} (\bibinfo {year} {2009})}\BibitemShut {NoStop}%
\bibitem [{\citenamefont {Segovia}\ \emph
  {et~al.}(2013{\natexlab{b}})\citenamefont {Segovia}, \citenamefont {Entem},\
  and\ \citenamefont {Fernandez}}]{Segovia:2013kg}%
  \BibitemOpen
  \bibfield  {author} {\bibinfo {author} {\bibfnamefont {J.}~\bibnamefont
  {Segovia}}, \bibinfo {author} {\bibfnamefont {D.~R.}\ \bibnamefont {Entem}},\
  and\ \bibinfo {author} {\bibfnamefont {F.}~\bibnamefont {Fernandez}},\
  }\bibfield  {title} {\bibinfo {title} {{Strong charmonium decays in a
  microscopic model}},\ }\href
  {https://doi.org/10.1016/j.nuclphysa.2013.07.004} {\bibfield  {journal}
  {\bibinfo  {journal} {Nucl. Phys. A}\ }\textbf {\bibinfo {volume} {915}},\
  \bibinfo {pages} {125} (\bibinfo {year} {2013}{\natexlab{b}})},\ \Eprint
  {https://arxiv.org/abs/1301.2592} {arXiv:1301.2592 [hep-ph]} \BibitemShut
  {NoStop}%
\bibitem [{\citenamefont {Segovia}\ \emph
  {et~al.}(2015{\natexlab{b}})\citenamefont {Segovia}, \citenamefont {Entem},\
  and\ \citenamefont {Fern\'andez}}]{Segovia:2014mca}%
  \BibitemOpen
  \bibfield  {author} {\bibinfo {author} {\bibfnamefont {J.}~\bibnamefont
  {Segovia}}, \bibinfo {author} {\bibfnamefont {D.~R.}\ \bibnamefont {Entem}},\
  and\ \bibinfo {author} {\bibfnamefont {F.}~\bibnamefont {Fern\'andez}},\
  }\bibfield  {title} {\bibinfo {title} {{Puzzles in hadronic transitions of
  heavy quarkonium with two pion emission}},\ }\href
  {https://doi.org/10.1103/PhysRevD.91.014002} {\bibfield  {journal} {\bibinfo
  {journal} {Phys. Rev. D}\ }\textbf {\bibinfo {volume} {91}},\ \bibinfo
  {pages} {014002} (\bibinfo {year} {2015}{\natexlab{b}})},\ \Eprint
  {https://arxiv.org/abs/1409.7079} {arXiv:1409.7079 [hep-ph]} \BibitemShut
  {NoStop}%
\bibitem [{\citenamefont {Segovia}\ \emph
  {et~al.}(2016{\natexlab{b}})\citenamefont {Segovia}, \citenamefont
  {Fernandez},\ and\ \citenamefont {Entem}}]{Segovia:2015raa}%
  \BibitemOpen
  \bibfield  {author} {\bibinfo {author} {\bibfnamefont {J.}~\bibnamefont
  {Segovia}}, \bibinfo {author} {\bibfnamefont {F.}~\bibnamefont {Fernandez}},\
  and\ \bibinfo {author} {\bibfnamefont {D.~R.}\ \bibnamefont {Entem}},\
  }\bibfield  {title} {\bibinfo {title} {{The Role of Spin-Flipping Terms in
  Hadronic Transitions of ${\Upsilon (4S)}$}},\ }\href
  {https://doi.org/10.1007/s00601-016-1063-7} {\bibfield  {journal} {\bibinfo
  {journal} {Few Body Syst.}\ }\textbf {\bibinfo {volume} {57}},\ \bibinfo
  {pages} {275} (\bibinfo {year} {2016}{\natexlab{b}})},\ \Eprint
  {https://arxiv.org/abs/1507.01607} {arXiv:1507.01607 [hep-ph]} \BibitemShut
  {NoStop}%
\bibitem [{\citenamefont {Mart\'\i{}n-Gonz\'alez}\ \emph
  {et~al.}(2022)\citenamefont {Mart\'\i{}n-Gonz\'alez}, \citenamefont {Ortega},
  \citenamefont {Entem}, \citenamefont {Fern\'andez},\ and\ \citenamefont
  {Segovia}}]{Martin-Gonzalez:2022qwd}%
  \BibitemOpen
  \bibfield  {author} {\bibinfo {author} {\bibfnamefont {B.}~\bibnamefont
  {Mart\'\i{}n-Gonz\'alez}}, \bibinfo {author} {\bibfnamefont {P.~G.}\
  \bibnamefont {Ortega}}, \bibinfo {author} {\bibfnamefont {D.~R.}\
  \bibnamefont {Entem}}, \bibinfo {author} {\bibfnamefont {F.}~\bibnamefont
  {Fern\'andez}},\ and\ \bibinfo {author} {\bibfnamefont {J.}~\bibnamefont
  {Segovia}},\ }\bibfield  {title} {\bibinfo {title} {{Toward the discovery of
  novel Bc states: Radiative and hadronic transitions}},\ }\href
  {https://doi.org/10.1103/PhysRevD.106.054009} {\bibfield  {journal} {\bibinfo
   {journal} {Phys. Rev. D}\ }\textbf {\bibinfo {volume} {106}},\ \bibinfo
  {pages} {054009} (\bibinfo {year} {2022})},\ \Eprint
  {https://arxiv.org/abs/2205.05950} {arXiv:2205.05950 [hep-ph]} \BibitemShut
  {NoStop}%
\bibitem [{\citenamefont {Segovia}\ \emph
  {et~al.}(2011{\natexlab{a}})\citenamefont {Segovia}, \citenamefont
  {Albertus}, \citenamefont {Entem}, \citenamefont {Fernandez}, \citenamefont
  {Hernandez},\ and\ \citenamefont {Perez-Garcia}}]{Segovia:2011dg}%
  \BibitemOpen
  \bibfield  {author} {\bibinfo {author} {\bibfnamefont {J.}~\bibnamefont
  {Segovia}}, \bibinfo {author} {\bibfnamefont {C.}~\bibnamefont {Albertus}},
  \bibinfo {author} {\bibfnamefont {D.~R.}\ \bibnamefont {Entem}}, \bibinfo
  {author} {\bibfnamefont {F.}~\bibnamefont {Fernandez}}, \bibinfo {author}
  {\bibfnamefont {E.}~\bibnamefont {Hernandez}},\ and\ \bibinfo {author}
  {\bibfnamefont {M.~A.}\ \bibnamefont {Perez-Garcia}},\ }\bibfield  {title}
  {\bibinfo {title} {{Semileptonic $B$ and $B_{s}$ decays into orbitally
  excited charmed mesons}},\ }\href
  {https://doi.org/10.1103/PhysRevD.84.094029} {\bibfield  {journal} {\bibinfo
  {journal} {Phys. Rev. D}\ }\textbf {\bibinfo {volume} {84}},\ \bibinfo
  {pages} {094029} (\bibinfo {year} {2011}{\natexlab{a}})},\ \Eprint
  {https://arxiv.org/abs/1107.4248} {arXiv:1107.4248 [hep-ph]} \BibitemShut
  {NoStop}%
\bibitem [{\citenamefont {Segovia}\ \emph
  {et~al.}(2011{\natexlab{b}})\citenamefont {Segovia}, \citenamefont {Entem},\
  and\ \citenamefont {Fernandez}}]{Segovia:2011zza}%
  \BibitemOpen
  \bibfield  {author} {\bibinfo {author} {\bibfnamefont {J.}~\bibnamefont
  {Segovia}}, \bibinfo {author} {\bibfnamefont {D.~R.}\ \bibnamefont {Entem}},\
  and\ \bibinfo {author} {\bibfnamefont {F.}~\bibnamefont {Fernandez}},\
  }\bibfield  {title} {\bibinfo {title} {{Charmonium resonances in e+ e-
  exclusive reactions around the psi(4415) region}},\ }\href
  {https://doi.org/10.1103/PhysRevD.83.114018} {\bibfield  {journal} {\bibinfo
  {journal} {Phys. Rev. D}\ }\textbf {\bibinfo {volume} {83}},\ \bibinfo
  {pages} {114018} (\bibinfo {year} {2011}{\natexlab{b}})}\BibitemShut
  {NoStop}%
\bibitem [{\citenamefont {Segovia}\ \emph
  {et~al.}(2012{\natexlab{b}})\citenamefont {Segovia}, \citenamefont
  {Albertus}, \citenamefont {Hernandez}, \citenamefont {Fernandez},\ and\
  \citenamefont {Entem}}]{Segovia:2012yh}%
  \BibitemOpen
  \bibfield  {author} {\bibinfo {author} {\bibfnamefont {J.}~\bibnamefont
  {Segovia}}, \bibinfo {author} {\bibfnamefont {C.}~\bibnamefont {Albertus}},
  \bibinfo {author} {\bibfnamefont {E.}~\bibnamefont {Hernandez}}, \bibinfo
  {author} {\bibfnamefont {F.}~\bibnamefont {Fernandez}},\ and\ \bibinfo
  {author} {\bibfnamefont {D.~R.}\ \bibnamefont {Entem}},\ }\bibfield  {title}
  {\bibinfo {title} {{Nonleptonic $B \to D^{(*)}D_{sJ}^{(*)}$ decays and the
  nature of the orbitally excited charmed-strange mesons}},\ }\href
  {https://doi.org/10.1103/PhysRevD.86.014010} {\bibfield  {journal} {\bibinfo
  {journal} {Phys. Rev. D}\ }\textbf {\bibinfo {volume} {86}},\ \bibinfo
  {pages} {014010} (\bibinfo {year} {2012}{\natexlab{b}})},\ \Eprint
  {https://arxiv.org/abs/1203.4362} {arXiv:1203.4362 [hep-ph]} \BibitemShut
  {NoStop}%
\bibitem [{\citenamefont {Ortega}\ \emph
  {et~al.}(2013{\natexlab{a}})\citenamefont {Ortega}, \citenamefont {Entem},\
  and\ \citenamefont {Fernandez}}]{Ortega:2012rs}%
  \BibitemOpen
  \bibfield  {author} {\bibinfo {author} {\bibfnamefont {P.~G.}\ \bibnamefont
  {Ortega}}, \bibinfo {author} {\bibfnamefont {D.~R.}\ \bibnamefont {Entem}},\
  and\ \bibinfo {author} {\bibfnamefont {F.}~\bibnamefont {Fernandez}},\
  }\bibfield  {title} {\bibinfo {title} {{Molecular Structures in Charmonium
  Spectrum: The $XYZ$ Puzzle}},\ }\href
  {https://doi.org/10.1088/0954-3899/40/6/065107} {\bibfield  {journal}
  {\bibinfo  {journal} {J. Phys. G}\ }\textbf {\bibinfo {volume} {40}},\
  \bibinfo {pages} {065107} (\bibinfo {year} {2013}{\natexlab{a}})},\ \Eprint
  {https://arxiv.org/abs/1205.1699} {arXiv:1205.1699 [hep-ph]} \BibitemShut
  {NoStop}%
\bibitem [{\citenamefont {Yang}\ \emph {et~al.}(2020)\citenamefont {Yang},
  \citenamefont {Ping},\ and\ \citenamefont {Segovia}}]{Yang:2020atz}%
  \BibitemOpen
  \bibfield  {author} {\bibinfo {author} {\bibfnamefont {G.}~\bibnamefont
  {Yang}}, \bibinfo {author} {\bibfnamefont {J.}~\bibnamefont {Ping}},\ and\
  \bibinfo {author} {\bibfnamefont {J.}~\bibnamefont {Segovia}},\ }\bibfield
  {title} {\bibinfo {title} {{Tetra- and penta-quark structures in the
  constituent quark model}},\ }\href {https://doi.org/10.3390/sym12111869}
  {\bibfield  {journal} {\bibinfo  {journal} {Symmetry}\ }\textbf {\bibinfo
  {volume} {12}},\ \bibinfo {pages} {1869} (\bibinfo {year} {2020})},\ \Eprint
  {https://arxiv.org/abs/2009.00238} {arXiv:2009.00238 [hep-ph]} \BibitemShut
  {NoStop}%
\bibitem [{\citenamefont {Ortega}\ \emph
  {et~al.}(2013{\natexlab{b}})\citenamefont {Ortega}, \citenamefont {Entem},\
  and\ \citenamefont {Fernandez}}]{Ortega:2012cx}%
  \BibitemOpen
  \bibfield  {author} {\bibinfo {author} {\bibfnamefont {P.~G.}\ \bibnamefont
  {Ortega}}, \bibinfo {author} {\bibfnamefont {D.~R.}\ \bibnamefont {Entem}},\
  and\ \bibinfo {author} {\bibfnamefont {F.}~\bibnamefont {Fernandez}},\
  }\bibfield  {title} {\bibinfo {title} {{Quark model description of the
  $\Lambda_c(2940)^+$ as a molecular $D^*N$ state and the possible existence of
  the$ \Lambda_b(6248)$}},\ }\href
  {https://doi.org/10.1016/j.physletb.2012.12.025} {\bibfield  {journal}
  {\bibinfo  {journal} {Phys. Lett. B}\ }\textbf {\bibinfo {volume} {718}},\
  \bibinfo {pages} {1381} (\bibinfo {year} {2013}{\natexlab{b}})},\ \Eprint
  {https://arxiv.org/abs/1210.2633} {arXiv:1210.2633 [hep-ph]} \BibitemShut
  {NoStop}%
\bibitem [{\citenamefont {Ortega}\ \emph
  {et~al.}(2014{\natexlab{a}})\citenamefont {Ortega}, \citenamefont {Entem},\
  and\ \citenamefont {Fern\'andez}}]{Ortega:2014fha}%
  \BibitemOpen
  \bibfield  {author} {\bibinfo {author} {\bibfnamefont {P.~G.}\ \bibnamefont
  {Ortega}}, \bibinfo {author} {\bibfnamefont {D.~R.}\ \bibnamefont {Entem}},\
  and\ \bibinfo {author} {\bibfnamefont {F.}~\bibnamefont {Fern\'andez}},\
  }\bibfield  {title} {\bibinfo {title} {{$D^*\Delta$ molecular interpretation
  for the $X_c$(3250)}},\ }\href
  {https://doi.org/10.1016/j.physletb.2013.12.058} {\bibfield  {journal}
  {\bibinfo  {journal} {Phys. Lett. B}\ }\textbf {\bibinfo {volume} {729}},\
  \bibinfo {pages} {24} (\bibinfo {year} {2014}{\natexlab{a}})}\BibitemShut
  {NoStop}%
\bibitem [{\citenamefont {Ortega}\ \emph
  {et~al.}(2014{\natexlab{b}})\citenamefont {Ortega}, \citenamefont {Entem},\
  and\ \citenamefont {Fern\'andez}}]{Ortega:2014eoa}%
  \BibitemOpen
  \bibfield  {author} {\bibinfo {author} {\bibfnamefont {P.~G.}\ \bibnamefont
  {Ortega}}, \bibinfo {author} {\bibfnamefont {D.~R.}\ \bibnamefont {Entem}},\
  and\ \bibinfo {author} {\bibfnamefont {F.}~\bibnamefont {Fern\'andez}},\
  }\bibfield  {title} {\bibinfo {title} {{Hadronic molecules in the open charm
  and open bottom baryon spectrum}},\ }\href
  {https://doi.org/10.1103/PhysRevD.90.114013} {\bibfield  {journal} {\bibinfo
  {journal} {Phys. Rev. D}\ }\textbf {\bibinfo {volume} {90}},\ \bibinfo
  {pages} {114013} (\bibinfo {year} {2014}{\natexlab{b}})}\BibitemShut
  {NoStop}%
\bibitem [{\citenamefont {Micu}(1969)}]{Micu:1968mk}%
  \BibitemOpen
  \bibfield  {author} {\bibinfo {author} {\bibfnamefont {L.}~\bibnamefont
  {Micu}},\ }\bibfield  {title} {\bibinfo {title} {{Decay rates of meson
  resonances in a quark model}},\ }\href
  {https://doi.org/10.1016/0550-3213(69)90039-X} {\bibfield  {journal}
  {\bibinfo  {journal} {Nucl. Phys. B}\ }\textbf {\bibinfo {volume} {10}},\
  \bibinfo {pages} {521} (\bibinfo {year} {1969})}\BibitemShut {NoStop}%
\bibitem [{\citenamefont {Le~Yaouanc}\ \emph {et~al.}(1973)\citenamefont
  {Le~Yaouanc}, \citenamefont {Oliver}, \citenamefont {Pene},\ and\
  \citenamefont {Raynal}}]{LeYaouanc:1972vsx}%
  \BibitemOpen
  \bibfield  {author} {\bibinfo {author} {\bibfnamefont {A.}~\bibnamefont
  {Le~Yaouanc}}, \bibinfo {author} {\bibfnamefont {L.}~\bibnamefont {Oliver}},
  \bibinfo {author} {\bibfnamefont {O.}~\bibnamefont {Pene}},\ and\ \bibinfo
  {author} {\bibfnamefont {J.~C.}\ \bibnamefont {Raynal}},\ }\bibfield  {title}
  {\bibinfo {title} {{Naive quark pair creation model of strong interaction
  vertices}},\ }\href {https://doi.org/10.1103/PhysRevD.8.2223} {\bibfield
  {journal} {\bibinfo  {journal} {Phys. Rev. D}\ }\textbf {\bibinfo {volume}
  {8}},\ \bibinfo {pages} {2223} (\bibinfo {year} {1973})}\BibitemShut
  {NoStop}%
\bibitem [{\citenamefont {Carlitz}\ and\ \citenamefont
  {Kislinger}(1970)}]{Carlitz:1970xb}%
  \BibitemOpen
  \bibfield  {author} {\bibinfo {author} {\bibfnamefont {R.~D.}\ \bibnamefont
  {Carlitz}}\ and\ \bibinfo {author} {\bibfnamefont {M.}~\bibnamefont
  {Kislinger}},\ }\bibfield  {title} {\bibinfo {title} {{Regge amplitude
  arising from su(6)w vertices}},\ }\href
  {https://doi.org/10.1103/PhysRevD.2.336} {\bibfield  {journal} {\bibinfo
  {journal} {Phys. Rev. D}\ }\textbf {\bibinfo {volume} {2}},\ \bibinfo {pages}
  {336} (\bibinfo {year} {1970})}\BibitemShut {NoStop}%
\bibitem [{\citenamefont {Le~Yaouanc}\ \emph {et~al.}(1974)\citenamefont
  {Le~Yaouanc}, \citenamefont {Oliver}, \citenamefont {Pene},\ and\
  \citenamefont {Raynal}}]{LeYaouanc:1973ldf}%
  \BibitemOpen
  \bibfield  {author} {\bibinfo {author} {\bibfnamefont {A.}~\bibnamefont
  {Le~Yaouanc}}, \bibinfo {author} {\bibfnamefont {L.}~\bibnamefont {Oliver}},
  \bibinfo {author} {\bibfnamefont {O.}~\bibnamefont {Pene}},\ and\ \bibinfo
  {author} {\bibfnamefont {J.~C.}\ \bibnamefont {Raynal}},\ }\bibfield  {title}
  {\bibinfo {title} {{Naive quark pair creation model and baryon decays}},\
  }\href {https://doi.org/10.1103/PhysRevD.9.1415} {\bibfield  {journal}
  {\bibinfo  {journal} {Phys. Rev. D}\ }\textbf {\bibinfo {volume} {9}},\
  \bibinfo {pages} {1415} (\bibinfo {year} {1974})}\BibitemShut {NoStop}%
\bibitem [{\citenamefont {Le~Yaouanc}\ \emph
  {et~al.}(1977{\natexlab{a}})\citenamefont {Le~Yaouanc}, \citenamefont
  {Oliver}, \citenamefont {Pene},\ and\ \citenamefont
  {Raynal}}]{LeYaouanc:1977fsz}%
  \BibitemOpen
  \bibfield  {author} {\bibinfo {author} {\bibfnamefont {A.}~\bibnamefont
  {Le~Yaouanc}}, \bibinfo {author} {\bibfnamefont {L.}~\bibnamefont {Oliver}},
  \bibinfo {author} {\bibfnamefont {O.}~\bibnamefont {Pene}},\ and\ \bibinfo
  {author} {\bibfnamefont {J.~C.}\ \bibnamefont {Raynal}},\ }\bibfield  {title}
  {\bibinfo {title} {{Strong Decays of psi-prime-prime (4.028) as a Radial
  Excitation of Charmonium}},\ }\href
  {https://doi.org/10.1016/0370-2693(77)90250-7} {\bibfield  {journal}
  {\bibinfo  {journal} {Phys. Lett. B}\ }\textbf {\bibinfo {volume} {71}},\
  \bibinfo {pages} {397} (\bibinfo {year} {1977}{\natexlab{a}})}\BibitemShut
  {NoStop}%
\bibitem [{\citenamefont {Le~Yaouanc}\ \emph
  {et~al.}(1977{\natexlab{b}})\citenamefont {Le~Yaouanc}, \citenamefont
  {Oliver}, \citenamefont {Pene},\ and\ \citenamefont
  {Raynal}}]{LeYaouanc:1977gm}%
  \BibitemOpen
  \bibfield  {author} {\bibinfo {author} {\bibfnamefont {A.}~\bibnamefont
  {Le~Yaouanc}}, \bibinfo {author} {\bibfnamefont {L.}~\bibnamefont {Oliver}},
  \bibinfo {author} {\bibfnamefont {O.}~\bibnamefont {Pene}},\ and\ \bibinfo
  {author} {\bibfnamefont {J.~C.}\ \bibnamefont {Raynal}},\ }\bibfield  {title}
  {\bibinfo {title} {{Why Is psi-prime-prime-prime (4.414) SO Narrow?}},\
  }\href {https://doi.org/10.1016/0370-2693(77)90062-4} {\bibfield  {journal}
  {\bibinfo  {journal} {Phys. Lett. B}\ }\textbf {\bibinfo {volume} {72}},\
  \bibinfo {pages} {57} (\bibinfo {year} {1977}{\natexlab{b}})}\BibitemShut
  {NoStop}%
\bibitem [{\citenamefont {Hayne}\ and\ \citenamefont
  {Isgur}(1982)}]{Hayne:1981zy}%
  \BibitemOpen
  \bibfield  {author} {\bibinfo {author} {\bibfnamefont {C.}~\bibnamefont
  {Hayne}}\ and\ \bibinfo {author} {\bibfnamefont {N.}~\bibnamefont {Isgur}},\
  }\bibfield  {title} {\bibinfo {title} {{Beyond the Wave Function at the
  Origin: Some Momentum Dependent Effects in the Nonrelativistic Quark
  Model}},\ }\href {https://doi.org/10.1103/PhysRevD.25.1944} {\bibfield
  {journal} {\bibinfo  {journal} {Phys. Rev. D}\ }\textbf {\bibinfo {volume}
  {25}},\ \bibinfo {pages} {1944} (\bibinfo {year} {1982})}\BibitemShut
  {NoStop}%
\bibitem [{\citenamefont {Blundell}\ and\ \citenamefont
  {Godfrey}(1996)}]{Blundell:1995ev}%
  \BibitemOpen
  \bibfield  {author} {\bibinfo {author} {\bibfnamefont {H.~G.}\ \bibnamefont
  {Blundell}}\ and\ \bibinfo {author} {\bibfnamefont {S.}~\bibnamefont
  {Godfrey}},\ }\bibfield  {title} {\bibinfo {title} {{The Xi (2220) revisited:
  Strong decays of the 1(3) F2 1(3) F4 s anti-s mesons}},\ }\href
  {https://doi.org/10.1103/PhysRevD.53.3700} {\bibfield  {journal} {\bibinfo
  {journal} {Phys. Rev. D}\ }\textbf {\bibinfo {volume} {53}},\ \bibinfo
  {pages} {3700} (\bibinfo {year} {1996})},\ \Eprint
  {https://arxiv.org/abs/hep-ph/9508264} {arXiv:hep-ph/9508264} \BibitemShut
  {NoStop}%
\bibitem [{\citenamefont {Guo}\ \emph {et~al.}(2019)\citenamefont {Guo},
  \citenamefont {Yang},\ and\ \citenamefont {Zhang}}]{Guo:2019ytq}%
  \BibitemOpen
  \bibfield  {author} {\bibinfo {author} {\bibfnamefont {J.-J.}\ \bibnamefont
  {Guo}}, \bibinfo {author} {\bibfnamefont {P.}~\bibnamefont {Yang}},\ and\
  \bibinfo {author} {\bibfnamefont {A.}~\bibnamefont {Zhang}},\ }\bibfield
  {title} {\bibinfo {title} {{Strong decays of observed $\Lambda_c$ baryons in
  the $^3P_0$ model}},\ }\href {https://doi.org/10.1103/PhysRevD.100.014001}
  {\bibfield  {journal} {\bibinfo  {journal} {Phys. Rev. D}\ }\textbf {\bibinfo
  {volume} {100}},\ \bibinfo {pages} {014001} (\bibinfo {year} {2019})},\
  \Eprint {https://arxiv.org/abs/1902.07488} {arXiv:1902.07488 [hep-ph]}
  \BibitemShut {NoStop}%
\bibitem [{\citenamefont {Kokoski}\ and\ \citenamefont
  {Isgur}(1987)}]{Kokoski:1985is}%
  \BibitemOpen
  \bibfield  {author} {\bibinfo {author} {\bibfnamefont {R.}~\bibnamefont
  {Kokoski}}\ and\ \bibinfo {author} {\bibfnamefont {N.}~\bibnamefont
  {Isgur}},\ }\bibfield  {title} {\bibinfo {title} {{Meson Decays by Flux Tube
  Breaking}},\ }\href {https://doi.org/10.1103/PhysRevD.35.907} {\bibfield
  {journal} {\bibinfo  {journal} {Phys. Rev. D}\ }\textbf {\bibinfo {volume}
  {35}},\ \bibinfo {pages} {907} (\bibinfo {year} {1987})}\BibitemShut
  {NoStop}%
\bibitem [{\citenamefont {Roberts}\ and\ \citenamefont
  {Silvestre-Brac}(1998)}]{Roberts:1997kq}%
  \BibitemOpen
  \bibfield  {author} {\bibinfo {author} {\bibfnamefont {W.}~\bibnamefont
  {Roberts}}\ and\ \bibinfo {author} {\bibfnamefont {B.}~\bibnamefont
  {Silvestre-Brac}},\ }\bibfield  {title} {\bibinfo {title} {{Meson decays in a
  quark model}},\ }\href {https://doi.org/10.1103/PhysRevD.57.1694} {\bibfield
  {journal} {\bibinfo  {journal} {Phys. Rev. D}\ }\textbf {\bibinfo {volume}
  {57}},\ \bibinfo {pages} {1694} (\bibinfo {year} {1998})},\ \Eprint
  {https://arxiv.org/abs/hep-ph/9708235} {arXiv:hep-ph/9708235} \BibitemShut
  {NoStop}%
\bibitem [{\citenamefont {Chen}\ \emph {et~al.}(2018)\citenamefont {Chen},
  \citenamefont {Ping}, \citenamefont {Roberts},\ and\ \citenamefont
  {Segovia}}]{Chen:2017mug}%
  \BibitemOpen
  \bibfield  {author} {\bibinfo {author} {\bibfnamefont {X.}~\bibnamefont
  {Chen}}, \bibinfo {author} {\bibfnamefont {J.}~\bibnamefont {Ping}}, \bibinfo
  {author} {\bibfnamefont {C.~D.}\ \bibnamefont {Roberts}},\ and\ \bibinfo
  {author} {\bibfnamefont {J.}~\bibnamefont {Segovia}},\ }\bibfield  {title}
  {\bibinfo {title} {{Light-meson masses in an unquenched quark model}},\
  }\href {https://doi.org/10.1103/PhysRevD.97.094016} {\bibfield  {journal}
  {\bibinfo  {journal} {Phys. Rev. D}\ }\textbf {\bibinfo {volume} {97}},\
  \bibinfo {pages} {094016} (\bibinfo {year} {2018})},\ \Eprint
  {https://arxiv.org/abs/1712.04457} {arXiv:1712.04457 [nucl-th]} \BibitemShut
  {NoStop}%
\bibitem [{\citenamefont {Okubo}(1963)}]{Okubo:1963fa}%
  \BibitemOpen
  \bibfield  {author} {\bibinfo {author} {\bibfnamefont {S.}~\bibnamefont
  {Okubo}},\ }\bibfield  {title} {\bibinfo {title} {{Phi meson and unitary
  symmetry model}},\ }\href {https://doi.org/10.1016/S0375-9601(63)92548-9}
  {\bibfield  {journal} {\bibinfo  {journal} {Phys. Lett.}\ }\textbf {\bibinfo
  {volume} {5}},\ \bibinfo {pages} {165} (\bibinfo {year} {1963})}\BibitemShut
  {NoStop}%
\bibitem [{\citenamefont {Zweig}(1964{\natexlab{b}})}]{Zweig:1964jf}%
  \BibitemOpen
  \bibfield  {author} {\bibinfo {author} {\bibfnamefont {G.}~\bibnamefont
  {Zweig}},\ }\bibinfo {title} {{An SU(3) model for strong interaction symmetry
  and its breaking. Version 2}},\ in\ \href@noop {} {\emph {\bibinfo
  {booktitle} {{DEVELOPMENTS IN THE QUARK THEORY OF HADRONS. VOL. 1. 1964 -
  1978}}}},\ \bibinfo {editor} {edited by\ \bibinfo {editor} {\bibfnamefont
  {D.~B.}\ \bibnamefont {Lichtenberg}}\ and\ \bibinfo {editor} {\bibfnamefont
  {S.~P.}\ \bibnamefont {Rosen}}}\ (\bibinfo {year} {1964})\ pp.\ \bibinfo
  {pages} {22--101}\BibitemShut {NoStop}%
\bibitem [{\citenamefont {Iizuka}(1966)}]{Iizuka:1966fk}%
  \BibitemOpen
  \bibfield  {author} {\bibinfo {author} {\bibfnamefont {J.}~\bibnamefont
  {Iizuka}},\ }\bibfield  {title} {\bibinfo {title} {{Systematics and
  phenomenology of meson family}},\ }\href {https://doi.org/10.1143/PTPS.37.21}
  {\bibfield  {journal} {\bibinfo  {journal} {Prog. Theor. Phys. Suppl.}\
  }\textbf {\bibinfo {volume} {37}},\ \bibinfo {pages} {21} (\bibinfo {year}
  {1966})}\BibitemShut {NoStop}%
\bibitem [{\citenamefont {Capstick}\ and\ \citenamefont
  {Roberts}(1993)}]{Capstick:1992th}%
  \BibitemOpen
  \bibfield  {author} {\bibinfo {author} {\bibfnamefont {S.}~\bibnamefont
  {Capstick}}\ and\ \bibinfo {author} {\bibfnamefont {W.}~\bibnamefont
  {Roberts}},\ }\bibfield  {title} {\bibinfo {title} {{N pi decays of baryons
  in a relativized model}},\ }\href {https://doi.org/10.1103/PhysRevD.47.1994}
  {\bibfield  {journal} {\bibinfo  {journal} {Phys. Rev. D}\ }\textbf {\bibinfo
  {volume} {47}},\ \bibinfo {pages} {1994} (\bibinfo {year}
  {1993})}\BibitemShut {NoStop}%
\bibitem [{\citenamefont {Varshalovich}\ \emph {et~al.}(1988)\citenamefont
  {Varshalovich}, \citenamefont {Moskalev},\ and\ \citenamefont
  {Khersonskii}}]{QTAM}%
  \BibitemOpen
  \bibfield  {author} {\bibinfo {author} {\bibfnamefont {D.~A.}\ \bibnamefont
  {Varshalovich}}, \bibinfo {author} {\bibfnamefont {A.~N.}\ \bibnamefont
  {Moskalev}},\ and\ \bibinfo {author} {\bibfnamefont {V.~K.}\ \bibnamefont
  {Khersonskii}},\ }\href@noop {} {\emph {\bibinfo {title} {{Quantum Theory of
  Angular Momentum}}}}\ (\bibinfo  {publisher} {World Scientific},\ \bibinfo
  {year} {1988})\BibitemShut {NoStop}%
\bibitem [{\citenamefont {Workman}\ and\ \citenamefont
  {Others}(2022)}]{Workman:2022ynf}%
  \BibitemOpen
  \bibfield  {author} {\bibinfo {author} {\bibfnamefont {R.~L.}\ \bibnamefont
  {Workman}}\ and\ \bibinfo {author} {\bibnamefont {Others}} (\bibinfo
  {collaboration} {Particle Data Group}),\ }\bibfield  {title} {\bibinfo
  {title} {{Review of Particle Physics}},\ }\href
  {https://doi.org/10.1093/ptep/ptac097} {\bibfield  {journal} {\bibinfo
  {journal} {PTEP}\ }\textbf {\bibinfo {volume} {2022}},\ \bibinfo {pages}
  {083C01} (\bibinfo {year} {2022})}\BibitemShut {NoStop}%
\bibitem [{\citenamefont {Straub}\ \emph {et~al.}(1988)\citenamefont {Straub},
  \citenamefont {Zhang}, \citenamefont {Brauer}, \citenamefont {Faessler},
  \citenamefont {Khadkikar},\ and\ \citenamefont {Lubeck}}]{Straub:1988gj}%
  \BibitemOpen
  \bibfield  {author} {\bibinfo {author} {\bibfnamefont {U.}~\bibnamefont
  {Straub}}, \bibinfo {author} {\bibfnamefont {Z.-Y.}\ \bibnamefont {Zhang}},
  \bibinfo {author} {\bibfnamefont {K.}~\bibnamefont {Brauer}}, \bibinfo
  {author} {\bibfnamefont {A.}~\bibnamefont {Faessler}}, \bibinfo {author}
  {\bibfnamefont {S.~B.}\ \bibnamefont {Khadkikar}},\ and\ \bibinfo {author}
  {\bibfnamefont {G.}~\bibnamefont {Lubeck}},\ }\bibfield  {title} {\bibinfo
  {title} {{Hyperon Nucleon Interaction in the Quark Cluster Model}},\ }\href
  {https://doi.org/10.1016/0375-9474(88)90092-9} {\bibfield  {journal}
  {\bibinfo  {journal} {Nucl. Phys. A}\ }\textbf {\bibinfo {volume} {483}},\
  \bibinfo {pages} {686} (\bibinfo {year} {1988})}\BibitemShut {NoStop}%
\end{thebibliography}%

\end{document}